%% file: Main_TAP_final.tex
\documentclass{IEEEtran}
\UseRawInputEncoding
\input{Z_Settings.tex}

\begin{document}

\title{All-Dielectric Ceramic 3-D Printed Unidirectional Leaky-Wave Antenna at mm-Wave Frequencies}
\author{Guillaume Fran\c{c}ois, \IEEEmembership{Member, IEEE}, Masoud Sakaki, Henrik Jansen, \\Amar Al-Bassam, \IEEEmembership{Member, IEEE}, Niels Benson, \IEEEmembership{Member, IEEE}, and Dirk Heberling, \IEEEmembership{Senior Member, IEEE}
\thanks{This work has been submitted to the IEEE for possible publication. Copyright may be transferred without notice, after which this version may no longer be accessible.}
\thanks{This work was funded by the German Federal Ministry of Education and Research (BMBF) in the course of the 6GEM research project under grant number 16KISK038. (\textit{Corresponding author: Guillaume Fran\c{c}ois.})}
\thanks{G. Fran\c{c}ois and H. Jansen are with the Institute of High Frequency Technology, RWTH Aachen University, 52074 Aachen, Germany (e-mail: francois@ihf.rwth-aachen.de).}
\thanks{M. Sakaki and N. Benson are with the Institute of Technology for Nanostructures, University of Duisburg-Essen, 47057 Duisburg, Germany.}
\thanks{A. Al-Bassam is with the Fraunhofer Institute for High Frequency Physics and Radar Techniques (FHR), 53343 Wachtberg, Germany.}
\thanks{D. Heberling is with the Institute of High Frequency Technology, RWTH Aachen University, 52074 Aachen, Germany, and also with the Fraunhofer Institute for High Frequency Physics and Radar Techniques (FHR), 53343 Wachtberg, Germany.}
}

\maketitle

\begin{abstract}

An all-dielectric leaky-wave antenna, radiating unidirectionally at broadside without the use of a metallic ground plane, is designed and manufactured. It operates in the \qtyrange{70}{90}{\GHz} frequency range with a design broadside frequency of \qty{77}{\GHz}. The design process makes use of the theory of photonic bandgaps to position a \textminus1 harmonic at the broadside frequency on the second Bragg-condition point. The structure is then modified to close the bandgap and to simultaneously guarantee maximum total field coupling to the relevant radiating harmonic, ensuring radiation unidirectionality. The antenna is manufactured with an \ce{Al2O3} ceramic printing process, demonstrating its manufacturing potential for complex low-loss high-frequency structures. Simulations agree well with measurements, with the main deviation being a \qty{1}{GHz} frequency shift that could be reproduced in simulation. The antenna prototype shows a scanning range of \qtyrange{-18}{22}{\degree} and gain values above \qty{23}{\dBi}. More than \qty{85}{\percent} of the radiation power is radiated unidirectionally at broadside, with simulations showing potential to attain \qty{98}{\percent}.  

\end{abstract}


\section{Introduction}\label{sec:Intro}
\IEEEPARstart{A}{dditive} manufacturing has become an increasingly appealing option for realizing radiating structures at mm-wave and sub-THz frequencies. Modern high-resolution printers can now reproduce fine geometrical features that satisfy the small size requirements for such frequencies. Dedicated materials with high-performance electrical properties are being developed for the available printing processes~\cite{Barrett_MOTL_05_2025}. When a comparatively high permittivity is required, ceramic-based materials have emerged as a preferred choice. Recent demonstrations include a dielectric resonator antenna at \qty{2}{\GHz} printed from \ce{TiO2}- and \ce{ZrO2}-based filaments by fused deposition modeling~\cite{Vandelle_EuCAP_05_2025}, a flat gradient-index lens at \qty{15}{\GHz} manufactured from a \ce{ZrO2} material by material jetting~\cite{Oh_APS_07_2020}, and a flat Luneburg lens at Ku-band produced from an \ce{MgTiO3} material by stereolithography~\cite{Lou_AWPL_02_2021}.

Beyond geometrical flexibility, 3\nobreakdash-D printed ceramics offer non-negligible electrical advantages. Pure or near-pure ceramics exhibit intrinsically low loss tangents, on the order of \num{e-4} over a wide frequency range~\cite{Di-Marco_JECS_11_2016}, roughly an order of magnitude below commercially available \ce{PTFE} ceramic-filled laminates~\cite{Carter_JIMT_10_2023}. Such low losses present a definite advantage for radiating structures operating at high-frequency, where the effects of substrate losses are more pronounced.

A common obstacle for 3\nobreakdash-D printed structures is the metallization required to realize conventional antenna topologies. A dual dielectric-metal printing process as in~\cite{Zhu_ISAP_11_2021} is an option, but it is however a complex process that must guarantee smooth metallic surface to avoid added surface-roughness losses. To the authors' knowledge no such process has yet been established for ceramic materials. Another possibility would be to metallize the printed structures, but this is an additional process with ongoing challenges~\cite{Romankov_MT_09_2025}. In both cases, even if an additional post-processing step, such as milling, can be used to smooth the metallic parts, it would only be valid for the exposed metallic surfaces and not the ones in contact with the ceramic interface, greatly reducing the effectiveness of the method for reducing surface roughness losses.

A more elegant solution is to completely eliminate the metallization  by designing radiating structures that require no metal, which consequently removes the conductor losses associated with skin depth and surface roughness at high frequencies. The remaining dielectric losses can in turn be kept very low through the use of ceramics. Such all-dielectric concepts are particularly attractive for radars operating at E-band and above, as well as for emerging communication and antenna-on-chip systems at sub-THz frequencies. In addition, the high thermal and mechanical robustness of ceramics makes them well suited to harsh operating environments, such as those involving high temperatures~\cite{Sharma_IEEEMM_08_2024}. Most of these systems rely on planar, highly directive antennas radiating in the transverse plane, a performance normally enabled by patch- metallic-grating-based leaky wave antennas (LWAs)~\cite{Klohn_TMTT_10_1978}. Periodic LWAs also provide frequency scanning~\cite{Oliner_Volakis_book_chap, Jackson_PIEEE_07_2012}, but may suffer from a broadside open stopband that degrades the antenna gain as the antenna scans through broadside. Recently, many  suppression approaches were established for metal-based LWAs, including unit-cell matching and the introduction of transverse asymmetry to the unit cell~\cite{Paulotto_TAP_07_2009, Otto_TAP_10_2014}.

However, removing the metal from such antennas is not straightforward. The metallic gratings of LWA can be replaced by dielectric gratings~\cite{Schwering_TMTT_02_1983}, but this topology still relies on a metallic ground plane to ensure unidirectionality (single-direction radiation). In the absence of the ground plane, the gratings scatter the guided wave into both the upper and lower half-spaces. Fully metal-free unidirectional radiators have been realized using tapered dielectric rods~\cite{Mueller_BSTJ_10_1947, Withayachumnankul_APLP_04_2018} or leakage from frustrated total internal reflection~\cite{Lees_APLP_03_2024}, yet these structures radiate predominantly in the endfire direction rather than at broadside.

A directly relevant solution originates from integrated optics, where unidirectional grating couplers for fiber-to-chip coupling have been studied extensively. Unidirectional grating couplers can direct \qtyrange{80}{90}{\percent} of the power into a single direction~\cite{Vermeulen_OE_08_2010, Bozzola_OE_06_2015}, while a dual-grating design achieves up to \qty{99}{\percent} unidirectionality with the added benefit of minimizing the reflection toward the source, thereby ensuring an excellent feed match~\cite{Michaels_OE_02_2018}.
Building on the optical principle in~\cite{Michaels_OE_02_2018} and its previous implementation using laser-engraved dielectric substrates~\cite{Francois_EuCAP_04_2026}, this paper presents a monolithic 3\nobreakdash-D-printed ceramic all-dielectric LWA (AD-LWA) operating in the frequency range \qtyrange{70}{90}{\GHz} with a broadside frequency of \qty{77}{\GHz}. The main contributions are fourfold.
\begin{enumerate}
    \item A simple analytical model based on array theory is presented to describe the scanning and unidirectional-radiation mechanisms of the shifted dual-grating AD-LWA.
    \item A Bloch/photonic bandgap (PBG)-based design procedure is presented, placing the $n=-1$ radiating space harmonic at broadside at the second-order Bragg point and identifies the corresponding open stopband at broadside. The grating parameters (dimensions and longitudinal shift) are adjusted to suppress the open stopband.
    \item A new implementation of a Mikaelian lens is presented to collimate the guided wave within the dielectric slab of the AD-LWA. The required permittivity profile is achieved through two complementary techniques, varying the profile height and the introduction of holes, in order to overcome the resolution limits of the printing process.
    \item A prototype is designed in pure alumina (\ce{Al2O3}), fabricated using stereolithography 3\nobreakdash-D printing, and later characterized. The prototype demonstrates that an efficient unidirectional radiation is compatible with a ceramic printing process at fine feature sizes.
\end{enumerate}

Following this introduction, the article is organized as follows. \secref{sec:Theory} presents the design methodology and theory of unidirectional dielectric radiating structures. We first present an analytical model based on array theory, which is then expanded with PBG theory. With it we refine the geometrical definitions of substrate thickness and unidirectionality condition, the PBG-based grating design, and the leaky-wave behavior obtained by closing the bandgap. \secref{sec:Proto} describes the prototype and its fabrication, including the 3\nobreakdash-D printing process, a Mikaelian lens design, and the final antenna. Next, \secref{sec:Res} presents the measurement setup and results, showing the reflection coefficient, radiation patterns and the unidirectionality performance. Finally, \secref{sec:conc} concludes the paper.  

\section{Radiating-structure Theory and Design}\label{sec:Theory}
\begin{figure}[!htbp]
    \centering
    \begin{subfigure}{\columnwidth}
        \centering
        \includegraphics[width=0.80\columnwidth]{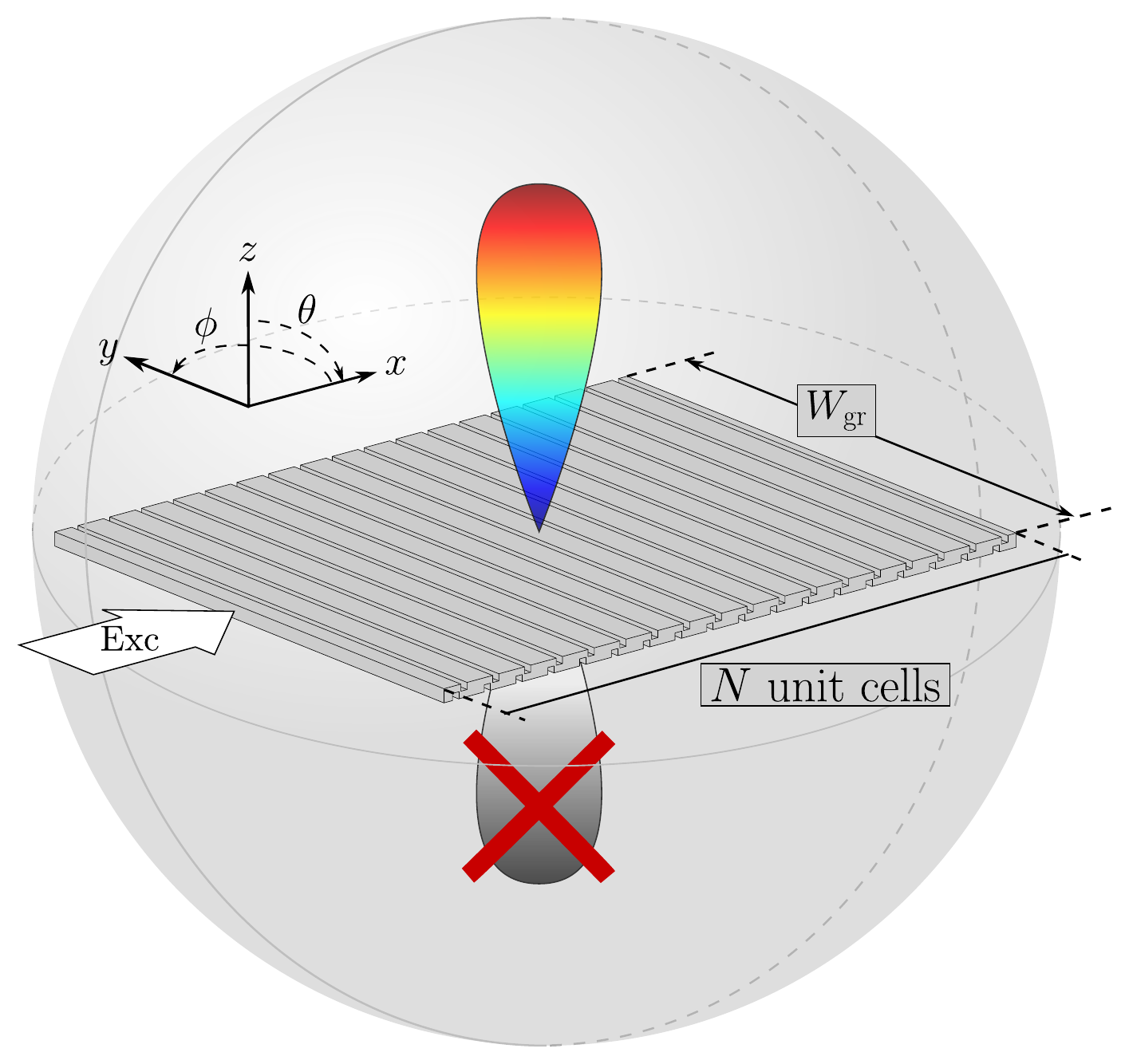}
        \caption{}
        \label{fig:dia_grat_3D}
    \end{subfigure}
    \begin{subfigure}[t]{0.49\columnwidth}
        \setlength\figurewidth{\linewidth}
        \setlength\figureheight{3.9cm}
        \centering
        \includegraphics[width=\linewidth]{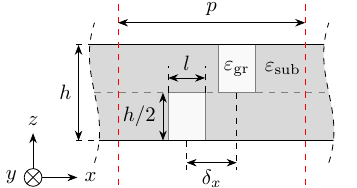}
        \caption{}
        \label{fig:dia_grat_geo}
    \end{subfigure}
    \hfill
    \begin{subfigure}[t]{0.49\columnwidth}
        \setlength\figurewidth{\linewidth}
        \setlength\figureheight{3.9cm}
        \centering
        \includegraphics[width=\linewidth]{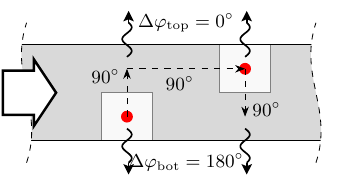}
        \caption{}
        \label{fig:dia_grat_interf}
    \end{subfigure}
		
    \caption{Radiating structures concept. \protect\subref{fig:dia_grat_3D}~Overview of the radiating structures, the unidirectional radiation and radiation/measurement sphere. \protect\subref{fig:dia_grat_geo}~Grating geometries of the AD-LWA and their corresponding parameters. The red dashed lines illustrate the ports positions and extensions for dispersion simulations described later. \protect\subref{fig:dia_grat_interf}~Zoomed-in visualization of the principle of interference for top and bottom directions to achieve radiation unidirectionality towards top only, with a wave excitation coming from the left.}
    \label{fig:dia_grat}
\end{figure}

The shifted dual-grating concept introduced in~\cite{Michaels_OE_02_2018} is adapted to realize the AD-LWA shown in Fig.~\ref{fig:dia_grat}\subref{fig:dia_grat_3D}. First, we obtain the phase and geometric conditions for the unidirectional radiation by employing a simple model based on array theory. Then, a Bloch-wave analysis based on photonic-bandgap theory is used to determine the substrate and grating parameters shown in Fig.~\ref{fig:dia_grat}\subref{fig:dia_grat_geo}, where the longitudinal grating shift $\delta_x$ is designed to suppress the open stopband at broadside while achieving unidirectional radiation. Finally, the AD-LWA with finite number of unit cells is simulated and its results are compared with the array model.

\subsection{Array model}\label{sec:ThArray}
To model the radiation mechanism of the AD-LWA and to obtain the related grating shift $\delta_x$ to achieve unidirectional radiation, we calculate first the radiation contribution from the shifted-grating pair of the unit cell and then of the full array. Fig~\ref{fig:dia_grat}\subref{fig:dia_grat_interf} illustrates a wave propagating from the left impinging on two point-like radiator elements, resembling the shifted gratings. These elements scatter the wave with a relative phase shift proportional to their relative position, and they interfere constructively ($\Delta\varphi_{\textrm{top}}=\qty{0}{\degree}$) when radiating towards the top and destructively ($\Delta\varphi_{\textrm{bot}}=\qty{180}{\degree}$) towards the bottom.\par
\begin{figure}
  \centering
  \begin{subfigure}{\linewidth}
    \centering
    \includegraphics[width=0.95\linewidth]{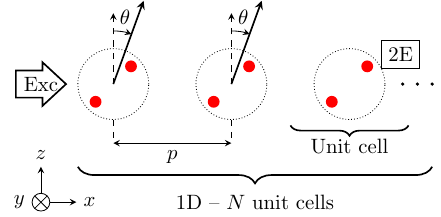}
    \caption{}
    \label{fig:dia_array_overview}
  \end{subfigure}
  \medskip
    \begin{subfigure}{0.49\linewidth}
    \centering
    \includegraphics[width=\linewidth]{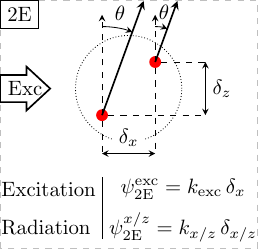}
    \caption{}
    \label{fig:dia_array_2e}
  \end{subfigure}
  \hfill
  \begin{subfigure}{0.49\linewidth}
    \centering
    \includegraphics[width=\linewidth]{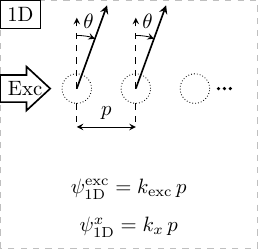}
    \caption{}
    \label{fig:dia_array_1d}
  \end{subfigure}

  \caption{Overview of the array model to demonstrate unidirectionality, and the different excitation and radiation phase shifts used in the calculation of the radiation patterns. \protect\subref{fig:dia_array_overview}~The total array configuration comprised of 1D and 2E radiation systems. \protect\subref{fig:dia_array_2e}~2E array geometry and geometric $\psi_{\textrm{2E}}$ phase shifts. \protect\subref{fig:dia_array_1d}~1D array geometry and geometric $\psi_{\textrm{1D}}$ phase shifts.}
  \label{fig:dia_array}
\end{figure}

In a more abstract way, the dual-grating pair of the unit cell is depicted in Fig.~\ref{fig:dia_array}\subref{fig:dia_array_overview}. The two point-like elements (labeled 2E) are placed with $x$ and $z$ offsets from each other, $\delta_x$ and $\delta_z$ [Fig.~\ref{fig:dia_array}\subref{fig:dia_grat_geo}], respectively. They are then repeated periodically $N\textrm{th}$ times with the period $p$ to construct a \mbox{1-D} array (labeled 1D). The \mbox{1-D} array is being excited by a propagating guided wave with a wavenumber $k_\textrm{exc}$ in the $+x$-direction [Fig.~\ref{fig:dia_array}\subref{fig:dia_array_2e}].\footnote{This simple \mbox{1-D} array model is sufficient to model the AD-LWA shown in Fig.~\ref{fig:dia_grat}\subref{fig:dia_grat_3D}, since all gratings are invariant in the $y$-direction.}

The total radiation pattern $\textrm{F}_{\textrm{T}}$ of this system is defined as~\cite{Balanis_AntennaTheory_Book_2016}:
\begin{equation}\label{eq:AF_total}
    \textrm{F}_{\textrm{T}} = \textrm{F}_{\textrm{2E}} \cdot \textrm{F}_{\textrm{1D}},
\end{equation}
\noindent effectively treating $\textrm{F}_{\textrm{2E}}$ as an element pattern multiplied by the array factor $\textrm{F}_{\textrm{1D}}$. Using~\cite{Balanis_AntennaTheory_Book_2016}, we may calculate the radiation patterns of $\textrm{F}_{\textrm{1D}}$ and $\textrm{F}_{\textrm{2E}}$ of the 1D and 2E systems, respectively, as:
\begin{subequations}\label{eq:AF_glob}
    \begin{gather}    
        \textrm{F}_{\textrm{2E}} = \left\vert 2 \cos \left( \frac{1}{2} \Psi_{\textrm{2E}} \right) \right\vert,
        \label{eq:AF_2E}\\
        \textrm{F}_{\textrm{1D}} = \left\vert \frac{\sin \left( \frac{N}{2} \Psi_{\textrm{1D}} \right)}{\sin \left( \frac{1}{2} \Psi_{\textrm{1D}} \right)} \right\vert,
        \label{eq:AF_1D}
    \end{gather}    
\end{subequations}
\noindent with $\Psi_{\textrm{2E}}$ and $\Psi_{\textrm{1D}}$ being the total relative phase shift between radiating elements in the 2E and 1D, respectively. They depend on the frequency $f$, the geometry of the system (relative position of elements), and the angle of radiation $\theta$. We separate the $\theta$ terms for a more simplified overview, and thus we have:
\begin{equation}\label{eq:AF_Psi_generic}
    \Psi = \sum_i \, \psi_i \left( f,d \right) \, g_i \left( \theta \right). 
\end{equation}
\noindent $\psi_i$ is the resulting phase shift from a wave propagation with wavenumber $k$ or $\beta$ (scattered/radiating or guided) over a specific distance $d$ ($p$ or $\delta_{x/z}$), thus in general $\psi = kd$. $g_i$ can be any function related to the angle of radiation with respect to the coordinate system. Applying~\eqref{eq:AF_Psi_generic} to the 2E and 1D systems we obtain:
\begin{subequations}\label{eq:AF_Psi_glob}
    \begin{gather}    
        \Psi_{\textrm{2E}} = \psi^{\textrm{exc}}_{\textrm{2E}} - \psi^x_{\textrm{2E}} \, \sin\theta - \psi^z_{\textrm{2E}} \, \cos\theta,
        \label{eq:AF_Psi_2E_glob}\\
        \Psi_{\textrm{1D}} = \psi^{\textrm{exc}}_{\textrm{1D}} - \psi^x_{\textrm{1D}} \, \sin\theta.
        \label{eq:AF_Psi_1D_glob}
    \end{gather}    
\end{subequations}
\noindent The $\psi_{\textrm{2E}}$ and $\psi_{\textrm{1D}}$ definitions are given in Fig.~\ref{fig:dia_array}\subref{fig:dia_array_2e} and~\subref{fig:dia_array_1d}, where the $f$ and $d$ dependencies are still included, but are dropped in \eqref{eq:AF_Psi_glob} for conciseness. The $\psi_{\textrm{exc}}$ terms correspond to the phase shifts generated by the propagating excitation wave in the system. The $\psi^{x/z}$ terms are linked to the scattering of the propagating wave, thus radiating and related to the angle of radiation $\theta$. The conditions in this model for the unidirectional radiation solely depend on the values $k$ and $d$. Thus, we propose two steps to obtain these values: 1) a simplified model giving an easy intuition on the possibility to obtain unidirectional radiation without considering the propagation properties of the guided wave within the dieletric slab, and 2) a model taking into account the realistic implementation of the AD-LWA.\par

\vspace{\baselineskip}
\noindent\textbf{Simplified analytical model:} The wave propagation and geometry influences on the phase shifts are both simplified. We consider that everywhere in the system the wave propagates with free-space characteristics, $k_{\textrm{exc}} = k_{x/z} = k_0$, resulting in $\psi = k_0d$ with $k_0 = 2\pi f/c$. For the 1\nobreakdash-D system, the terms in~\eqref{eq:AF_Psi_1D_glob} simplify to:
\begin{equation}\label{eq:AF_Psi_1D_glob_simp}
    \Psi^{\textrm{simp}}_{\textrm{1D}} = \frac{2\pi f}{c} p \left( 1 - \sin\theta \right).
\end{equation}
\noindent The goal is to guarantee maximum radiation at broadside, which is achieved when taking the maximum of \eqref{eq:AF_1D} to be $\Psi^{\textrm{simp}}_{\textrm{1D}} = 2\pi n$ ($n \in \mathbb{N}$) at $f = f_{\textrm{bs}}$ and $\theta=0$. Since the excitation term cannot be 0, we have the general condition necessary for maximum broadside radiation with $n=1$: 
\begin{equation}\label{eq:AF_1D_max}
    \Psi_{\textrm{1D}} = 2\pi.
\end{equation}
\noindent By inserting \eqref{eq:AF_Psi_1D_glob_simp} into \eqref{eq:AF_1D_max} at $f_{\textrm{bs}}$, we directly derive the two geometrical relationships:
\begin{subequations}\label{eq:AF_1D_simpl_geom_cond}
    \begin{gather}    
        k_0^{\textrm{bs}} \, p = 2\pi,
        \label{eq:AF_1D_simpl_geom_cond_p2pi}\\
        f_{\textrm{bs}} = c/p.
        \label{eq:AF_1D_simpl_geom_cond_pfbs}
    \end{gather}    
\end{subequations}
\noindent We then obtain the total radiation pattern $\textrm{F}^{\textrm{simp}}_{\textrm{1D}}$ by inserting \eqref{eq:AF_1D_simpl_geom_cond_pfbs} into \eqref{eq:AF_1D} as:
\begin{equation}\label{eq:AF_1D_simp}
    \textrm{F}^{\textrm{simp}}_{\textrm{1D}} =  \left\vert \frac{\sin \left( N\pi \frac{f}{f_{\textrm{bs}}} \left( 1 - \sin\theta \right) \right)}{\sin \left( \pi \frac{f}{f_{\textrm{bs}}} \left( 1 - \sin\theta \right) \right)} \right\vert.
\end{equation}
Next, we analyze the 2E system by first assuming $\delta = \delta_x = \delta_z$, which upon insertion in~\eqref{eq:AF_Psi_2E_glob} yields the simplified phase shift expression in the 2E system $\Psi^{\textrm{simp}}_{\textrm{2E}}$:
\begin{equation}\label{eq:AF_Psi_2E_glob_simp}
    \Psi^{\textrm{simp}}_{\textrm{2E}} = \frac{2\pi f}{c} \delta \left( 1 - \sin\theta - \cos\theta \right).
\end{equation}
\noindent Since the goal is to achieve unidirectional radiation, the 2E radiation pattern has to cancel in the backward direction from broadside, i.e., $\textrm{F}_{\textrm{2E}} = 0$ at $f_{\textrm{bs}}$ and $\theta=\pi$, which from \eqref{eq:AF_2E} gives the condition:
\begin{equation}\label{eq:AF_2E_null}
    \Psi_{\textrm{2E}} = \pi.
\end{equation}
\noindent After inserting \eqref{eq:AF_Psi_2E_glob_simp} into \eqref{eq:AF_2E_null}, and with the previous relationship in~\eqref{eq:AF_1D_simpl_geom_cond_pfbs}, we derive the geometrical shift:
\begin{equation}\label{eq:AF_2E_simpl_geom_cond_pdelta}
    \delta = \delta_x = \delta_z = \frac{p}{4} = \frac{\lambda_0^{\mathrm{bs}}}{4}.
\end{equation}
This relation shows that the period $p$ in the 1\nobreakdash-D array and the geometrical offset $\delta$ between the gratings pair in the 2E system are both related to the wavelength of the propagating excitation wave at broadside. We remind that we consider here the simplified model, thus we obtain $\lambda_0^{\mathrm{bs}}$ as a free-space wavenumber, but later we will have to consider the real guided wave within the dielectric slab.\\
We finally obtain the radiation pattern $\textrm{F}^{\textrm{simp}}_{\textrm{2E}}$ by inserting \eqref{eq:AF_2E_null} and \eqref{eq:AF_2E_simpl_geom_cond_pdelta} into \eqref{eq:AF_2E} as:
\begin{equation}\label{eq:AF_2E_simp}
    \textrm{F}^{\textrm{simp}}_{\textrm{2E}} = \left\vert 2 \cos \left( \pi \frac{f}{4f_{\textrm{bs}}} \left( 1 - \sin\theta - \cos\theta \right)  \right) \right\vert.
\end{equation}
We verify the expected unidirectional radiation by numerically calculating the individual radiation patterns for the 1D (with $N=18$ unit cells) and 2E systems, given in \eqref{eq:AF_1D_simp} and \eqref{eq:AF_2E_simp}, respectively; as well as the total field evaluated by~\eqref{eq:AF_total}, with results shown in Fig.~\ref{fig:F_simp}. Furthermore, the patterns are calculated for three frequencies around the broadside frequency, as $f_{\textrm{bs}}-\Delta f$, $f_{\textrm{bs}}$ and $f_{\textrm{bs}}+\Delta f$ ($\Delta f/f_{\textrm{bs}}=\qty{13}{\percent}$) to show the unidirectional radiation behavior as the main beam scans through the broadside direction. It can clearly be shown that the 2E patterns [Fig.~\ref{fig:F_simp}\subref{fig:F_2E_simp}] cancel in the backward direction, while backlobes of the radiation patterns are visible for the 1-D array [Fig.~\ref{fig:F_simp}\subref{fig:F_1D_simp}]. Upon multiplication of both patterns, the backlobe in the total radiation pattern, $\textrm{F}^{\textrm{simp}}_{\textrm{T}}$, is clearly canceled at $f_{\textrm{bs}}$ or otherwise reduced, as shown in Fig.~\ref{fig:F_simp}\subref{fig:F_T_simp}.\footnote{Due to the simplified assumptions, the wave characteristics are all considered in free space, with a clear resulting consequence of having high grating lobes from the 1\nobreakdash-D array  [Fig.~\ref{fig:F_simp}\subref{fig:F_1D_simp}] occurring at $\theta=\qty{90}{\degree}$ as well as other angles for frequencies higher than $f_{\textrm{bs}}$. This is simply due to the element spacing $p=\lambda_0$, which will guarantee the appearance of grating lobes~\cite{Balanis_AntennaTheory_Book_2016}.}

\newlength{\twoEDwidth}
\newlength{\Twidth}
\setlength{\twoEDwidth}{3.6cm}   
\setlength{\Twidth}{5.8cm}       

\begin{figure}[!htbp]
  \centering
  \begin{minipage}[c]{\twoEDwidth}
    \centering
    \captionsetup[subfigure]{skip=-8pt}
    \begin{subfigure}{\linewidth}
      \centering
      \includegraphics[width=\linewidth]{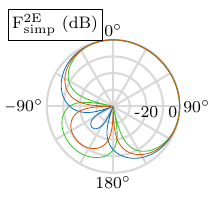}
      \caption{}
      \label{fig:F_2E_simp}
    \end{subfigure}

    \begin{subfigure}{\linewidth}
      \centering
      \includegraphics[width=\linewidth]{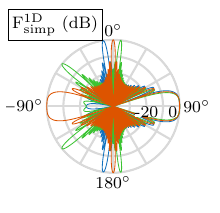}
      \caption{}
      \label{fig:F_1D_simp}
    \end{subfigure}
  \end{minipage}
  \hspace{-0.8cm}
  \captionsetup[subfigure]{skip=-13pt}
  \begin{subfigure}[c]{\Twidth}
    \centering
    \includegraphics[width=\linewidth]{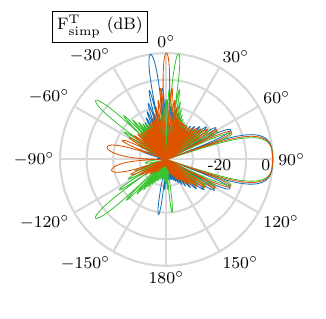}
    \caption{}
    \label{fig:F_T_simp}
  \end{subfigure}
  \par\vspace{0.5em}
  \centering
  \input{Tikz/F_simp_legend.tikz}
  \caption{Radiation patterns of the simplified array model elements for three frequencies ($\Delta f/f_{\textrm{bs}}=\qty{13}{\percent}$). \protect\subref{fig:F_2E_simp} $\textrm{F}^{\textrm{simp}}_{\textrm{2E}}$ where the back radiation canceling can be observed. \protect\subref{fig:F_1D_simp} $\textrm{F}^{\textrm{simp}}_{\textrm{1D}}$ with $N=18$ elements. \protect\subref{fig:F_T_simp} $\textrm{F}^{\textrm{simp}}_{\textrm{T}}$ as the total radiation pattern, with extremely low back radiation, especially at $f_{\textrm{bs}}$.}
  \label{fig:F_simp}
\end{figure}


\vspace{\baselineskip}
\noindent\textbf{Full analytical model:} To accurately model the radiation from the AD-LWA, we have to take into consideration the physical properties of the propagating wave with the dielectric slab. The excitation wave is propagating as a guided wave within the substrate in which the point-like radiators are embedded, as illustrated in Fig.~\ref{fig:dia_grat}\subref{fig:dia_grat_interf}, with their relative geometrical offsets given in Fig.~\ref{fig:dia_array}\subref{fig:dia_array_2e}. We first list all the definitions of this model:

\begin{outline}
    \1 The slow wave excitation is defined as a $\textrm{TE}_{0}$ guided wave in a dielectric waveguide (DWG) with wavenumber $\beta > k_0$.
    \1 The geometrical offsets of 2E are distinct, i.e. $\delta_x \neq \delta_z$.
    \1 The phase shifts $\psi^x_{\textrm{2E}}$ and $\psi^z_{\textrm{2E}}$ representing the scattered wave are taking into account the substrate in which the waves scatter, thus $\psi^x_{\textrm{2E}}=k_{\textrm{sub}} \, \delta_x$ and $\psi^z_{\textrm{2E}}=k_{\textrm{sub}} \, \delta_z$, with $k_{\textrm{sub}} = k_0 \, \sqrt{\varepsilon_{\textrm{sub}}}$ the substrate's wavenumber. Notice that these are not guided waves and hence not related to $\beta$.
\end{outline}

\noindent Inserting these conditions into \eqref{eq:AF_Psi_glob} yields the following phase shift definitions:
\begin{subequations}\label{eq:AF_Psi_glob_real}
    \begin{gather}   
        \Psi_{\textrm{2E}} = \beta\delta_x - k_{\textrm{sub}} \, \delta_x \, \sin\theta - k_{\textrm{sub}} \,\delta_z \, \cos\theta,
        \label{eq:AF_Psi_2E_glob_real}\\
        \Psi_{\textrm{1D}} = \beta p - k_0 \, p \sin\theta.
        \label{eq:AF_Psi_1D_glob_real}
    \end{gather}    
\end{subequations}
\noindent Similarly to the simplified analytical model [\eqref{eq:AF_1D_max} and \eqref{eq:AF_2E_null}], we can calculate the phase-shift conditions to achieve the required radiation pattern. Again, we start with the 1\nobreakdash-D array condition for maximum broadside radiation at $f_{\textrm{bs}}$ by inserting~\eqref{eq:AF_Psi_1D_glob_real} into \eqref{eq:AF_1D_max} with $\theta=0$ to obtain the relationship:
\begin{equation}\label{eq:AF_1D_real_geom_cond_p2pi}
    \beta_{\textrm{bs}} \, p = 2\pi,
\end{equation}
\noindent which is closely related to \eqref{eq:AF_1D_simpl_geom_cond_p2pi}, but taking into account the guided-wave characteristic of the excitation. 
Next, we investigate the 2E system and since it is geometrically more complex, we must define its geometry by implementing two radiation conditions at $f_{\textrm{bs}}$: maximum radiation at $\theta=0$ and null at $\theta=\pi$. For the maximum condition, we set:
\begin{equation}\label{eq:AF_2E_max}
    \Psi_{\textrm{2E}} = 0,
\end{equation}
\noindent which we insert into \eqref{eq:AF_Psi_2E_glob_real}. Thus at $f_{\textrm{bs}}$ and $\theta=0$, we deduce that:
\begin{equation}\label{eq:AF_2E_real_geom_cond_deltax}
    \delta_x = \frac{k_{\textrm{sub}}^{\textrm{bs}}}{\beta_{\textrm{bs}}}\delta_z,
\end{equation}
which gives the condition for the geometrical shift between the gratings in the $x$-direction ($\delta_x$). Using the 2E null condition given in \eqref{eq:AF_2E_null} and using \eqref{eq:AF_2E_real_geom_cond_deltax}, we obtain the final relationships:
\begin{equation}\label{eq:AF_2E_real_geom_cond_deltaz}
    k_{\textrm{sub}}^{\textrm{bs}} \, \delta_z = \frac{\pi}{2},
\end{equation}
\noindent and thus consequently from~\eqref{eq:AF_2E_real_geom_cond_deltax} we obtain also 
\begin{equation}\label{eq:AF_2E_real_geom_cond_betadeltax}
    \beta_{\textrm{bs}} \, \delta_x = \frac{\pi}{2}.
\end{equation}
\noindent These simple geometrical relationships linking all the phase shifts in the $x$ and $z$ directions to \qty{90}{\degree} are consistent with the relations in~\cite{Michaels_OE_02_2018}, but are related here clearly to the guided and substrate wavenumbers of the guided waves. Additionally, we have shown the required relation between the period and the guided wavelength of the excitation. To evaluate this analytical model, we need to derive the expressions for the wavenumbers, $\beta$ and $k_\textrm{sub}$, and relate them to the geometrical parameters, as will be presented next. 
\subsection{Photonic bandgap and space-harmonic design}\label{sec:ThOpenBG}
To obtain the conditions for unidirectional radiation with a TE\textsubscript{0} slow wave guiding system, the steps employed will be as follows:
\begin{outline}
    \1 We first define the substrate permittivity and height which are required to obtain the guided wavelength characteristics of a TE\textsubscript{0} mode in a DWG.
    \1 We will then change the DWG to a PBG in order to generate harmonics in the substrate, of which the $n=-1$ harmonic will later be radiating in a controlled way. The geometrical characteristics of the PBG needed to generate these harmonics with the correct space and frequency characteristics are then given.
    \1 After the harmonics are generated, we close the bandgap to allow the harmonics to propagate in the AD-LWA. At the same time, the structure should allow the harmonic $n=-1$ only to radiate, in one direction. Later in \secref{sec:ThCloseBG} we describe the simulation methods used to optimize the AD-LWA geometry, obtaining
    both the conditions of harmonic propagation (bandgap closing) and unidirectional radiation present for one geometry.
\end{outline}

The design process starts by fixing the substrate's permittivity, which is here restricted to the printing material, alumina (\ce{Al2O3}), with relative permittivity value $\varepsilon_{\textrm{sub}} = 9.2$ defined using the material characterization given in~\cite{Jenkel_IEEEA_10_2025}. The gratings themselves are taken as vacuum with $\varepsilon_{\textrm{gr}} = 1$. The operating broadside frequency of the antenna will be $f_{\textrm{bs}} = \qty{77}{\GHz}$, as was mentioned in \secref{sec:Intro}.\par

First, we obtain the vertical geometrical shift between the gratings ($\delta_z$) by using \eqref{eq:AF_2E_real_geom_cond_deltaz} and :
\begin{equation}\label{eq:deltaz_lambda}
    \delta_z = \dfrac{\pi}{2k_\textrm{sub}^\textrm{bs}} = \frac{\lambda_{\textrm{sub}}^{\textrm{bs}}}{4}.
\end{equation}
\noindent This relationship provides the required vertical shift to satisfy the condition in \eqref{eq:AF_2E_real_geom_cond_deltaz}, but it does not give the condition for the substrate height. For the substrate height, an air-dielectric interface matching condition has to be satisfied, and it reads~\cite{Pozar_MicroEng_Book_2012}:
\begin{equation}\label{eq:PBG_height}
    h = \frac{\lambda_{\textrm{sub}}^{\textrm{bs}}}{2} = \frac{c_0}{2 f_\textrm{bs}  \sqrt{\varepsilon_{\textrm{sub}}}},
\end{equation}
\noindent hence, the distance between the centers of the gratings in the $z$-direction (modeled as point-like radiators in \secref{sec:ThArray}) [Fig.~\ref{fig:dia_grat}\subref{fig:dia_grat_interf}] is kept as $\lambda_{\textrm{sub}}^{\textrm{bs}}/4$, as required by \eqref{eq:deltaz_lambda}.

\begin{figure}[!htbp]
    \centering
    \includegraphics[width=\columnwidth]{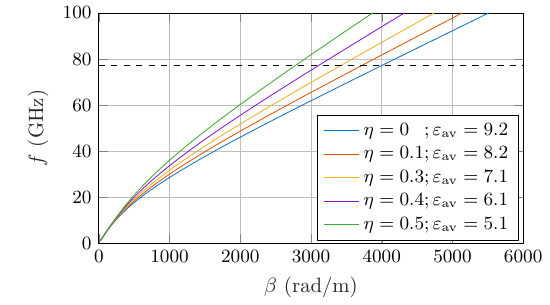}
    \caption{Dispersion diagrams of the TE\textsubscript{0} modes in a DWG of height $h$ and different $\varepsilon_{\textrm{av}}$ ($\eta$ of the gratings in the range [0; 0.5]). At $f_{\textrm{bs}} = \qty{77}{\GHz}$, the $\beta$ values correspond to our design second-order Bragg condition.}
    \label{fig:BetaTE0FF}
\end{figure}

\begin{figure}[!htbp]
    \centering
    \includegraphics[width=\columnwidth]{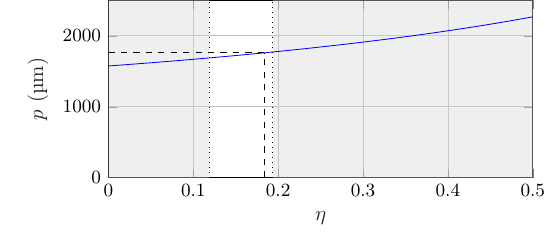}
    \caption{Relationship between $p$ and $\eta$ obtained from \eqref{eq:ChEqFinal} (plotted in Fig.~\ref{fig:BetaTE0FF}) and \eqref{eq:Period}. Manufacturing constraints require minimum values for $l$ and $\delta_x$ which forbid some values of $\eta$ - these are illustrated as gray areas in the figure. The selected period $p=\qty{1760}{\um}$ and its corresponding $\eta$ value are shown in dashed lines.} 
    \label{fig:pFF}
\end{figure}

After establishing the parameters for the DWG, we introduce the PBG for harmonic generation, which is related to the presense of the gratings. A PBG results from a periodic modulation of a medium permittivity~\cite{Peng_TMTT_01_1975, Yeh_SNY_06_2008}, generated in our case by the successive arrangement of the gratings in the substrate. We fist consider the case $\delta_x=\qty{0}{\um}$, a classical PBG structure as shown in Fig.~\ref{fig:dia_grat}\subref{fig:dia_grat_geo}. We can therefore define the average relative permittivity
\begin{equation}\label{eq:eps_av}
    \varepsilon_{\textrm{av}} = (1-\eta)\cdot\varepsilon_{\textrm{sub}}+\eta\cdot\varepsilon_{\textrm{gr}},
\end{equation}
\noindent as the weighted average relative permittivity of the PBG, with $\eta = l/p$ the fill factor of the gratings relative to the unit cell period. The PBG will be designed with $f_{\textrm{bs}}$ at the second-order Bragg coupling, where the $n=-1$ harmonic is in the radiating region~\cite{Jaggard_JOSA_07_1976}. This implies that the wavelength $\lambda$ at an $m_\textrm{th}$ order Bragg condition is related to the period $p$ by the relation $\lambda=2p/m$. In our case, $m=2$ and $\lambda=\lambda_{\textrm{g},\varepsilon_{\textrm{av}}}^{\textrm{bs}}$, corresponding to having the guided TE\textsubscript{0} wavelength of the broadside frequency (center of the bandgap) at the second Bragg condition in the average (unperturbed) permittivity medium. This yields the relationship
\begin{equation}\label{eq:Period}
    p=\lambda_{\textrm{g},\varepsilon_{\textrm{av}}}^{\textrm{bs}},
\end{equation}
\noindent which is a similar condition as given in \eqref{eq:AF_1D_real_geom_cond_p2pi}, when using the general relationship $\lambda_{\textrm{g}}^{\textrm{bs}}=2\pi/\beta_{\textrm{bs}}$. To obtain the TE\textsubscript{0} wavenumber of a guided wave in a standard symmetric DWG with average relative permittivity $\varepsilon_{\textrm{av}}$, we can use the even mode formula as in~\cite{Marcuse_AP_XX_1991}:
\begin{equation}\label{eq:ChEqGen}
    \kappa_{\textrm{in}} \cdot \mathrm{tan}\left(\kappa_{\textrm{in}}\frac{h}{2}\right) = \kappa_{\textrm{out}},
\end{equation}
\noindent with $\kappa_{\textrm{in}}$ and $\kappa_{\textrm{out}}$ the transverse wavenumbers inside and outside (air medium) the slab respectively. Using the relationships
\begin{subequations}
    \begin{gather}    
        \kappa_{\textrm{in}}=\sqrt{k_\textrm{0}^2 \cdot \varepsilon_{\textrm{av}}-\beta^2},
        \label{eq:Kappa}\\
        \kappa_{\textrm{out}}=\sqrt{\beta^2 - k_\textrm{0}^2},
        \label{eq:Gamma}
    \end{gather}    
\end{subequations}
\noindent where $k_\textrm{0}$ and $\beta$ are the free-space and DWG guided wavenumbers, respectively. Then, rearranging \eqref{eq:Kappa} to substitute $\beta$ in \eqref{eq:Gamma}, we may obtain the final characteristic equation with respect to $\kappa_{\textrm{in}}$:
\begin{equation}\label{eq:ChEqFinal}
    \kappa_{\textrm{in}} \cdot \mathrm{tan}\left(\kappa_{\textrm{in}}\frac{h}{2}\right) = \sqrt{k_\textrm{0}^2 \cdot (\varepsilon_{\textrm{av}} - 1) - \kappa_{\textrm{in}}^2}.
\end{equation}
By solving \eqref{eq:ChEqFinal} and using \eqref{eq:Kappa}, we obtain the wavenumber $\beta$ of the TE\textsubscript{0} mode. Fig.~\ref{fig:BetaTE0FF} shows the dispersion diagrams for different fill factors $\eta$ in the range [0; 0.5], where $\eta=0$ corresponds to the extreme case of pure substrate without gratings. At $f_{\textrm{bs}} = \qty{77}{\GHz}$, the $\beta_{\textrm{bs}}$ values correspond to our design second-order Bragg condition. With $\lambda_{\textrm{g},\varepsilon_{\textrm{av}}}^{\textrm{bs}}=2\pi/\beta_{\textrm{bs}}$ and \eqref{eq:Period} we can then relate the required period $p$ to $\eta$ as shown in Fig.~\ref{fig:pFF}.

\begin{figure}[!htbp]
    \centering
    \setlength\figureheight{2.6cm}
    \newlength{\panelwidth}
    \settowidth{\panelwidth}{\includegraphics[height=\figureheight]{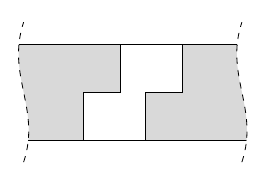}}

    \begin{subfigure}[t]{\panelwidth}
        \centering
        \includegraphics[height=\figureheight]{PDF/Dia_GratingsBad.pdf}
        \caption{}
        \label{fig:dia_grat_bad}
    \end{subfigure}
    \hspace{4mm}
    \begin{subfigure}[t]{\panelwidth}
        \centering
        \includegraphics[height=\figureheight]{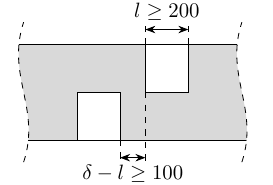}
        \caption{}
        \label{fig:dia_geo_req}
    \end{subfigure}
    \caption{AD-LWA grating geometries related to printing considerations. \protect\subref{fig:dia_grat_bad}~Incompatible grating configuration since it creates discontinued substrate parts. \protect\subref{fig:dia_geo_req}~Minimum requirements for $l$ and $\delta_x$ due to print quality and mechanical robustness considerations. These impact the achievable $\eta$ and thus $\varepsilon_{\textrm{av}}$.}
    \label{fig:dia_geo}
\end{figure}

The appropriate combination of $p$ and $\eta$ is next selected to obtain the final PBG configuration. This choice is strongly influenced by two practical considerations about the gratings manufacturing feasibility:

\begin{enumerate}
    \item  The grating length $l$ needs to have a minimum value of \qty{200}{\um}, otherwise manufacturing quality will be suboptimal (e.g. gratings filled with printing material instead of being hollow).
    \item We require a horizontal shift between the gratings pair by $\delta_x=\lambda_{\textrm{g},\varepsilon_{\textrm{av}}}^{\textrm{bs}}/4=p/4$, as derived from \eqref{eq:AF_2E_real_geom_cond_betadeltax} and \eqref{eq:Period}. It is thus important to avoid a configuration in which the gratings create a substrate discontinuity, as this would lead to disjointed parts of the PBG, as illustrated in Fig.~\ref{fig:dia_geo}\subref{fig:dia_grat_bad}. Therefore, we require a minimum \qty{100}{\um} distance between the two gratings, which forces the relationship $\delta_x-l \geq 100$ and consequently gives $p \geq \frac{100}{0.25-\eta}$. This condition is achieved for $\eta \leq 0.194$. These two points are illustrated in Fig.~\ref{fig:dia_geo}\subref{fig:dia_geo_req}.
\end{enumerate}
Hence, for the final design we select $p=\qty{1760}{\um}$ as it gives a large $l$, lowering the risk of obtaining gratings filled with printing material, while ensuring overall mechanical robustness with enough material between the gratings. We therefore obtain $\eta=0.183$, resulting in $\varepsilon_{\textrm{av}}=7.7$, $l=\qty{322}{\um}$ and $\delta_x=\qty{440}{\um}$.

\begin{figure}[!htbp]
    \centering
    \begin{tikzpicture}
      \node (mainplot) {\includegraphics[width=\columnwidth]{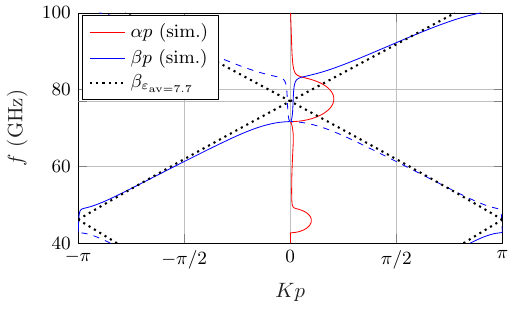}};
      \node[anchor=east, inner sep=1pt]
          at ([xshift=-0.037\columnwidth, yshift=-0.19168\columnwidth]mainplot.north east)
          {\includegraphics[width=0.2\columnwidth]{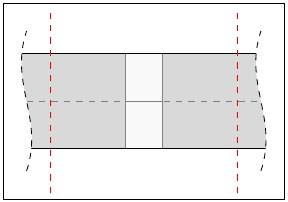}};
    \end{tikzpicture}
    \caption{Reduced Brillouin zone showing the dispersion diagram of the PBG as defined previously (here $\delta_x=\qty{0}{\um}$). The folded wavenumber $\beta_{\varepsilon_\textrm{av}=7.7}$ of a TE\textsubscript{0} mode of a DWG of height $h$ and $\varepsilon_{\textrm{av}}=7.7$ (corresponding to the grating configuration) is also shown. One can observe the second-order Bragg condition $Kp=0$ at $f_{\textrm{bs}} = \qty{77}{\GHz}$ which defines the center of the bandgap.}
    \label{fig:PBGTh}
\end{figure}

Fig.~\ref{fig:PBGTh} shows the resulting dispersion diagram of this grating configuration simulated in CST Microwave Studio. The simulation was performed on one unit cell, as shown in Fig.~\ref{fig:dia_grat}\subref{fig:dia_grat_geo}, where the ports were applied on the $yz$-plane of the substrate and extending \qty{2}{\mm} in the $z$ direction above and below the PBG; open boundaries were set in the $z$ direction to obtain the TE\textsubscript{0} mode behaviour of the PBG. The Bloch wavenumber $K$ is then extracted from the calculated two-port network parameters (ABCD)~\cite{Pozar_MicroEng_Book_2012}, as $Kp =\acos\left[(A+D)/2\right]$, with $K = \beta -j \alpha$, and where $\beta$ and $\alpha$ are the propagation and attenuation constants, respectively. The dispersion diagram is displayed in the reduced Brillouin zone $Kp=[-\pi; \pi]$ in the frequency range \qtyrange{40}{100}{\GHz}. For $\delta_x=\qty{0}{\um}$, the second bandgap appearing in the frequency range \qtyrange{71.7}{83.3}{\GHz} is clearly visible, especially when compared to the wavenumber $\pm \beta_{\varepsilon_\textrm{av}=7.7}$ of the unperturbed TE\textsubscript{0} mode of a DWG of height $h$ and $\varepsilon_{\textrm{av}}=7.7$. Here, we can observe the second-order Bragg condition $Kp=0$ at $f_{\textrm{bs}} = \qty{77}{\GHz}$, which defines the center of the bandgap.

\subsection{Unidirectional leaky-wave mechanism}\label{sec:ThCloseBG}
\begin{figure}[!htbp]
    \centering
    \begin{tikzpicture}
      \node (mainplot) {\includegraphics[width=\columnwidth]{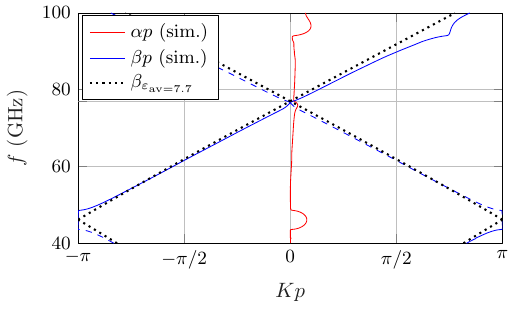}};
      \node[anchor=east, inner sep=1pt]
          at ([xshift=-0.037\columnwidth, yshift=-0.19168\columnwidth]mainplot.north east)
          {\includegraphics[width=0.2\columnwidth]{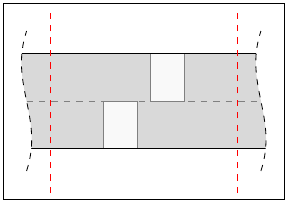}};
    \end{tikzpicture}
    \caption{Reduced Brillouin zone showing the dispersion diagram of the AD-LWA with $\delta_x=\qty{440}{\um}$ as defined in \secref{sec:ThOpenBG}. One can observe the bandgap closing at the second-order Bragg condition $Kp=0$ at $f_{\textrm{bs}} = \qty{77}{\GHz}$.}
    \label{fig:ClosedPBGTh}
\end{figure}

After we defined the bandgap structure to obtain the broadside frequency $f_{\textrm{bs}}$ at the second Bragg condition ($K=0$), we close the bandgap as a next step, which will allow the wave to propagate and leak the energy in one preferred direction [Fig.~\ref{fig:dia_grat}\subref{fig:dia_grat_3D}]. Following~\cite{Michaels_OE_02_2018}, the grating is "sliced" horizontally and the two halves are shifted by $\delta_x$. There are three phenomena than can be observed:
\begin{enumerate}
    \item\label{ite:LWA_ph_closeBGP} Closing the bandgap: In this configuration, the two harmonics ($n=\pm 1$ at the second Bragg condition) that were previously coupling to form standing waves in the periodic structure do not couple anymore. They exist as degenerate modes that travel independently in the PBG. In this case we would observe a reduced attenuation $\alpha$ in the bandgap zone around broadside, and a phase propagation constant $\beta$ converging to the $\beta_{\varepsilon_{\textrm{av}}}$ of the unperturbed average DWG.
    
    \item\label{ite:LWA_ph_energy} Maximal field coupling: A plane wave impinging from the top on the structure needs to couple optimally in the AD-LWA. At the optimal design geometries we would observe a maximum field intensity in the AD-LWA.
    
    \item\label{ite:LWA_ph_unidir} Unidirectional radiation: The propagating harmonic radiates outside of the structure only in one direction (top or bottom). In this case, we expect a plane wave impinging on the structure from the top (or bottom) to couple to a specific harmonic only ($n=+1$ or $-1$). We would then observe a strong difference between the field intensity of each harmonic.
\end{enumerate}
The first point (closing the bandgap) is obtained by implementing $\delta_x=\qty{440}{\um}$, as defined in \secref{sec:ThOpenBG}. Fig.~\ref{fig:ClosedPBGTh} shows the simulated dispersion diagram of the AD-LWA under this condition, where it can be seen that the $n=\pm 1$ harmonics are now almost converged to the behaviour of the guided wavenumber of the unperturbed TE\textsubscript{0}. We can also observe a reduced attenuation $\alpha$, the remainder being primarily a consequence of energy loss through radiation.

\begin{figure}[!htbp]
    \centering
    \includegraphics[width=\columnwidth]{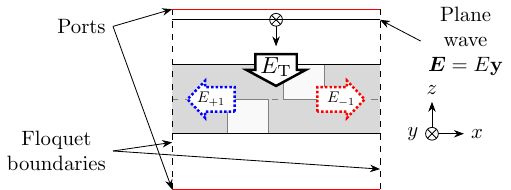}
    \caption{Illustration of the simulation setup to obtain field intensities in the AD-LWA structure with an E-field monitor configured in the structure. A plane wave generated on top of the structure impinges on it. Total and harmonic field intensities ($\lvert E_\mathrm{T} \rvert ^2$ and $\lvert E_{\pm 1} \rvert^2$ respectively) are then computed. In the optimal grating configuration, the plane wave propagating in the $-z$ direction should generate the maximum total and $n=+1$ harmonic intensities in the structure.}
    \label{fig:dia_dm_rev}
\end{figure}

To observe the phenomena in the points \ref{ite:LWA_ph_energy} and \ref{ite:LWA_ph_unidir}, a different simulation setup is considered in CST, shown in Fig.~\ref{fig:dia_dm_rev}. Here, we define a plane wave excitation at the broadside direction propagating in the normal direction towards the AD-LWA ($-z$). In this configuration, the excitation from the top should imply a dominant $+1$ harmonic propagating in the $-x$ direction, as illustrated in Fig.~\ref{fig:dia_grat}\subref{fig:dia_grat_interf}.\footnote{The sign inversion of the space harmonic arises from time-reversal symmetry, reflecting the directional duality between the receive and transmit configurations.} 

\begin{figure*}[t]
		\centering
        \input{Tikz/Field_harmonics_legend.tikz} 
		\begin{subfigure}{0.48\linewidth}
			\includegraphics[width=\linewidth]{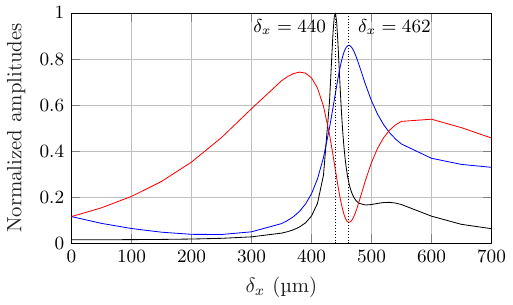}
			\caption{}
			\label{fig:har_norm_th}
		\end{subfigure}
		\begin{subfigure}{0.48\linewidth}
			\includegraphics[width=\linewidth]{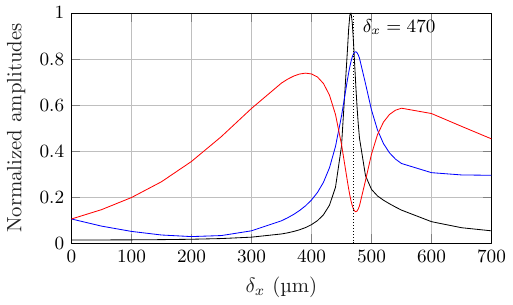}
			\caption{}
			\label{fig:har_norm_AD-LWA}
		\end{subfigure}
		\caption{Total field intensity coupled to the structure, and the $n=\pm1$ harmonics field intensities - both are normalized. \protect\subref{fig:har_norm_th}~Structure based on the theoretical considerations of~\secref{sec:ThOpenBG} with $l=\qty{322}{\um}$. \protect\subref{fig:har_norm_AD-LWA}~Structure obtained after parametric investigations, $l=\qty{346}{\um}$. The period is $p=\qty{1760}{\um}$ for both cases. One can observe a better alignment of the total field coupling and the maximum of the $n=+1$ harmonic in~\subref{fig:har_norm_AD-LWA} for $\delta_x=\qty{470}{\um}$ --- in this case we have close to maximum power coupling in the structure, transferred unidirectionally in the preferred $-x$ direction as illustrated in Fig.~\ref{fig:dia_dm_rev}.}
		\label{fig:harmonics_norm}
\end{figure*}

For point \ref{ite:LWA_ph_energy}, we monitor the E-field in the structure at $f_{\textrm{bs}}$ in the $xz$-plane within the radiating structure at $y=0$, and compute the total average field intensity, which contains all space harmonics contributions. This field intensity, $\lvert E_\mathrm{T} \rvert ^2$, is defined as:
\begin{equation}\label{eq:field_int_tot}
    \lvert E_\mathrm{T} \rvert ^2 = \frac{1}{h} \frac{1}{p} \int_0^h \int_0^p |E(x,z)|^2\,dx\,dz.
\end{equation}
For point \ref{ite:LWA_ph_unidir}, we compute the average field intensity $\lvert E_n\rvert ^2$ of the $n = \pm1$ harmonics only. We employ Bloch's theorem to calculate the E-field of the propagating wave~\cite{Yeh_SNY_06_2008}:
\begin{equation}\label{eq:field_Bloch}
    E(x,z)=\sum_{n=-\infty}^{\infty} E_n(z) e^{-j(K+2\pi n/p)x},
\end{equation}
\noindent with $E(x,z)$ and $K$ being the E-field monitored in simulation and the Bloch wavenumber in the AD-LWA, respectively. Hence, we may obtain the amplitudes of the space harmonics $E_n(z)$ by applying the Fourier series along one period $p$ of the AD-LWA. This can be written as:
\begin{equation}\label{eq:field_harmonic_x}
    E_n(z) = \frac{1}{p} \int_0^p E(x,z) e^{j (2\pi n/p)x}\,dx,
\end{equation}
\noindent where $K=0$ since we are considering the broadside case. Then, we compute the harmonic field intensity and average it over the entire height $h$ of the structure. The full average in 2\nobreakdash-D is:
\begin{equation}\label{eq:field_harmonic_int}
    \lvert E_n \rvert ^2 = \frac{1}{h} \int_{0}^{h} \lvert E_n(z) \rvert ^2 \,dz.
\end{equation}
\noindent Fig.~\ref{fig:harmonics_norm}\subref{fig:har_norm_th} plots the normalized total field intensity $\lvert E_\mathrm{T} \rvert ^2 / \mathrm{max}(|E_\mathrm{T}|^2)$, corresponding to the field at each $\delta_x$ normalized with respect to the maximum field intensity found in the $\delta_x$ investigation range. Additionally, we plot the normalized intensities $\lvert E_{\pm 1} \rvert^2 \,_\mathrm{Norm.} = \lvert E_{\pm 1} \rvert ^2 /\lvert E_\mathrm{T} \rvert ^2$, corresponding to the $n=\pm 1$ harmonics intensities normalized with respect to the total field intensity. The simulation is repeated for different horizontal shifts $\delta_x$.\par
We observe that the average field intensity $\lvert E_\mathrm{T} \rvert ^2$ has a maxima at $\delta_x=\qty{440}{\um}$, meaning that the incoming field from the plane wave couples efficiently in the structure at this point. However, the $n=\pm1$ harmonics intensities are not at their strongest difference: the maximum difference between them occurs at a different $\delta_x=\qty{462}{\um}$ value. Therefore, we can conclude that although the structure can couple the incoming wave for some $\delta_x$, it may not discriminate efficiently between the $\pm1$ harmonics and hence, would lead to a degraded unidirectional radiation, due the presence of a standing wave.\par
The mismatch between the field coupling and the harmonics can further be illustrated by Fig.~\ref{fig:ET_comp_har}. The top plots show the normalized E-field profile along $x$ at a specific height $z=h/2$ (the middle of the structure), broken in its real and imaginary parts. The bottom plots show the normalized magnitudes of $\lvert E_n \rvert_{\textrm{Norm.}}=\lvert E_n \rvert / \lvert E_\mathrm{T} \rvert$ of the Fourier series coefficients\footnote{We remind the relationship $\lvert E_\mathrm{T}(z) \rvert ^2 = \sum_{n=-\infty}^{+\infty}{\lvert E_n(z) \rvert ^2}$ relating the total field to its harmonics.} (truncated at $n=\pm3$) of this E-field profile. Figs.~\ref{fig:ET_comp_har}\subref{fig:ET_comp_har_d440} and \ref{fig:ET_comp_har}\subref{fig:ET_comp_har_d462} plot the field characteristics for $\delta_x=\qty{440}{}$ and $\qty{462}{\um}$, respectively. The field profile of \subref{fig:ET_comp_har_d462} shows a lower amplitude, but the phase difference between its real and imaginary parts is higher, nearing the characteristic field profile of a pure $n=+1$ harmonic (of the form $e^{jkx}$). This is further demonstrated in the bottom plots, where the $+1$ harmonic is much more dominant for the $\qty{462}{\um}$ case.\par

\newlength{\Eheight}
\setlength{\Eheight}{7.01444cm}

\begin{figure}[!htbp]
    \centering
    \captionsetup[subfigure]{oneside,margin={2cm,0cm}} 
    \begin{subfigure}{0.44\linewidth}
        \centering
			\includegraphics[height=\Eheight]{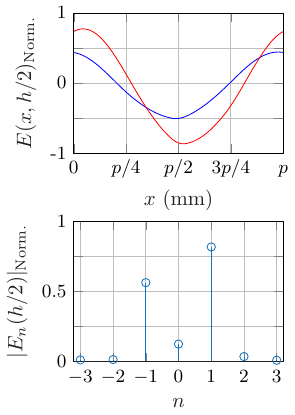}
			\caption{}
			\label{fig:ET_comp_har_d440}
    \end{subfigure}
    \hfill
    \captionsetup[subfigure]{oneside,margin={0cm,0cm}} 
    \begin{subfigure}{0.44\linewidth}
        \centering
        \includegraphics[height=\Eheight]{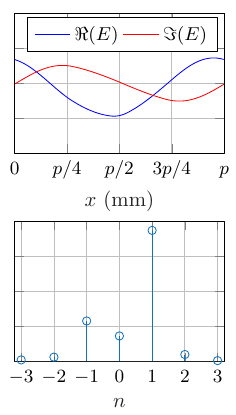}
        \caption{}
        \label{fig:ET_comp_har_d462}
    \end{subfigure}
    \caption{Illustration of the normalized E-field profile and its harmonics along one period at the height $z=h/2$ for the structure given in Fig.~\ref{fig:harmonics_norm}\subref{fig:har_norm_th}. \protect\subref{fig:ET_comp_har_d440}~Configuration with $\delta_x=\qty{440}{\um}$, where the field in the structure is at its maximum, but the $n=\pm1$ harmonics are not at their highest amplitude difference, meaning we have a reduced unidirectionality. \protect\subref{fig:ET_comp_har_d462}~$\delta_x=\qty{462}{\um}$ - in this case the unidirectionality is at its highest, but with a worse total energy coupling.}
    \label{fig:ET_comp_har}
\end{figure}

The first design steps thus deliver a geometry which is not yet perfect for the purposes of bandgap closing, field coupling and unidirectionality occurring optimally all at the same time. This discrepancy can be explained by the fact that the simple geometric relationships for unidirectionality derived in~\secref{sec:ThOpenBG} did not take into account all harmonics couplings. Indirect harmonics coupling occurs in periodic structures even between higher space harmonics, which can impact the behavior of both the bandgap closing and radiation, as stated in \cite{Jaggard_JOSA_07_1976, Lee_PRR_10_2025}. A comprehensive investigation of this type of coupling is beyond the scope of this paper. Instead, an alternative AD-LWA geometry is obtained from a parametric study, with its results presented in Fig.~\ref{fig:harmonics_norm}\subref{fig:har_norm_AD-LWA}. Compared with the idealized geometries, two parameters were modified: $l=\qty{346}{\um}$ and $\delta_x=\qty{470}{\um}$. The maxima of $\lvert E_\mathrm{T}\rvert^2$ and $\lvert E_{+1}\rvert^2$ are more closely aligned for this geometry. Among the values considered, $\delta_x=\qty{470}{\um}$ provides the best compromise between a strong desired-field response and sufficient harmonic separation, and hence would support a pure traveling-wave propagation.\par

\subsection{Validation results}\label{sec:ThSimRes}

In this section, we finalize the design process of the radiating structures by presenting and comparing the radiation characteristics the full analytical model presented in \secref{sec:ThArray} with a full-wave simulation obtained by CST:

\begin{outline}
    \1 For the analytical model in \secref{sec:ThArray}, all the geometrical parameters will follow the theoretical relationships given in~\eqref{eq:AF_1D_real_geom_cond_p2pi},~\eqref{eq:AF_2E_real_geom_cond_deltaz} and~\eqref{eq:AF_2E_real_geom_cond_betadeltax}. Nevertheless, the $\beta$ variation with frequency cannot be given theoretically, as any wavenumber in a DWG results from a numerical solution without a closed form expression. All $\beta$ values used in the pattern formulas will therefore derive from the theoretical PBG structure in~\secref{sec:ThOpenBG}, as $h$, $\varepsilon_{\textrm{av}}=7.7$, $p$, $l=\qty{322}{\um}$ and $\delta_x=\qty{440}{\um}$. the radiation pattern is then defined as $\textrm{F}_{\textrm{T}}$ with the conditions given in~\secref{sec:ThArray}.\footnote{In the analytical array-factor calculations, leakage-induced amplitude decay along the aperture is neglected by setting $\alpha=0$. This approximation does not alter the main-beam angle, although it only affects the beamwidth and sidelobe levels.}

    \1 The radiating geometry developed in~\secref{sec:ThCloseBG}, which was selected to provide strong field coupling while maintaining sufficient separation between the $n=\pm1$ spatial harmonics. The geometrical parameters are $h$, $\varepsilon_{\textrm{av}}=7.7$, $p$, $l=\qty{346}{\um}$ and $\delta_x=\qty{470}{\um}$. The results are obtained from CST full-wave simulations with matched ports at both ends of the grating section. Thus, any residual guided power reaching the output port is absorbed, allowing the intrinsic radiation behavior to be evaluated without reflections from the termination.
\end{outline}

\noindent To represent the structures intended for fabrication, presented in the forthcoming section, both radiating models comprise $N=18$ unit cells. The CST full-wave model also includes the finite extent along the $y$-direction; therefore, the simulated radiation results correspond to a grating width of $W_\textrm{gr}=\qty{26.4}{\mm}$, as illustrated in Fig.~\ref{fig:dia_grat}\subref{fig:dia_grat_3D}.\footnote{The selected values of $N$ and $W_\textrm{gr}$ are determined by manufacturing constraints, as will be discussed in the following \secref{sec:Proto}.}\par

\begin{figure}[!htbp]
    \centering
    \includegraphics[width=0.8\columnwidth]{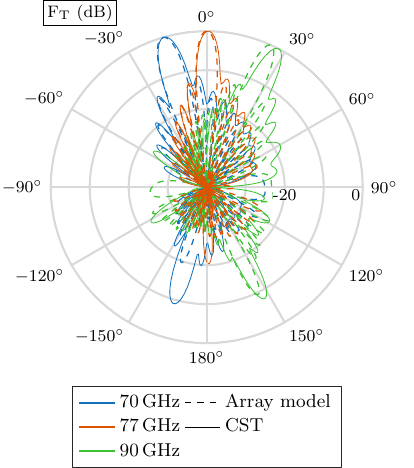}
    \caption{Normalized radiation patterns $\textrm{F}_{\textrm{T}}$ of the array model described in~\secref{sec:ThArray}, and the real radiating structures defined in~\secref{sec:ThCloseBG} and simulated in CST. A good agreement is observed concerning the scanning angles, especially for lower frequencies. There is a \qty{20}{\dB} radiation difference between the top and back directions at broadside frequency, confirming good unidirectionality performance.}
    \label{fig:F_T_CST_combined}
\end{figure}

The normalized radiation patterns of these models are shown in Fig.~\ref{fig:F_T_CST_combined}, over the frequency points 70, 77 and \qty{90}{\GHz}. We can observe the scanning property of the structures, with the array model matching quite well the simulation. Due to the slow wave implementation of the guided wave, $\beta > k_0$ and hence $p<\lambda_0$, there are no strong grating lobes as was the case in Fig.~\ref{fig:F_simp}. At $f_{\textrm{bs}}$, we can see that the array model shows very low back radiation when compared to the slightly worse performance obtained from the simulation, with about \qty{-20}{\dB} difference between the top and bottom. These considerations are also valid for the other two frequency points. Additionally, the agreement between the scanning angles predicted by the array model and the full-wave simulation deteriorates at \qty{90}{\GHz}. This discrepancy arises because the array model assumes a pure TE\textsubscript{0} excitation, whose propagation constant deviates more strongly from the corresponding Bloch wavenumber at this frequency, as shown in Fig.~\ref{fig:ClosedPBGTh}. 

To quantify and compare the radiation directionality of the two models, the harmonic-field analysis used in \secref{sec:ThCloseBG} is not applicable because the analytical array model does not provide the internal electric-field quantities required for that analysis. We therefore introduce a far-field metric that can be evaluated consistently for the analytical model, full-wave simulations, and subsequent measurements. The antenna is intended to concentrate its radiation in a single half-space, corresponding to the upper hemisphere ($z>0$) in Fig.~\ref{fig:dia_grat}\subref{fig:dia_grat_3D}. Accordingly, the upper-hemisphere radiated-power ratio $R$ is defined as the power radiated into this hemisphere normalized by the total power radiated over the full sphere: 
\begin{equation}\label{eq:unidir_R_CST}
R = \frac{\int_0^{2\pi} \int_0^{\frac{\pi}{2}} U \sin \theta \, d\theta \, d\phi}{\int_0^{2\pi} \int_0^\pi U \sin \theta \, d\theta \, d\phi},
\end{equation}
\noindent where $U$ is the radiation intensity, $\phi$ the angle from the $+x$-axis towards $+y$, and $\theta$ the angle from the $+z$-axis towards $+x$. \par

The definition of \eqref{eq:unidir_R_CST} applies to the finite CST models, for which the radiated power is integrated over the full sphere. In contrast, the analytical array model is translationally invariant along the $y$-direction and therefore represents a two-dimensional radiation problem in the $xz$-plane. The radiated power per unit length is thus evaluated in cylindrical coordinates by integrating the angular power density over a circle rather than a sphere. Accordingly, $R$ is obtained by normalizing the power radiated into the upper semicircle by that radiated over the full circle:
\begin{equation}\label{eq:unidir_R_array}
    R = \frac{\int_{\rvert \theta \lvert \leq \frac{\pi}{2}} \lvert \textrm{F}_{\textrm{T}} \vert^2 \, d\theta}{\int_{\rvert \theta \lvert \leq \pi} {\lvert \textrm{F}_{\textrm{T}} \vert^2 \, d\theta}},
\end{equation}
\noindent where $\rvert \theta \lvert \leq \pi/2$ corresponds to the upper semicircle.\footnote{Notice also the $\sin \theta$ of~\eqref{eq:unidir_R_CST} disappearing since we use cylindrical coordinates and drop the $\phi$ parameter.}\par

\begin{figure}[!htbp]
    \centering
    \includegraphics[width=\columnwidth]{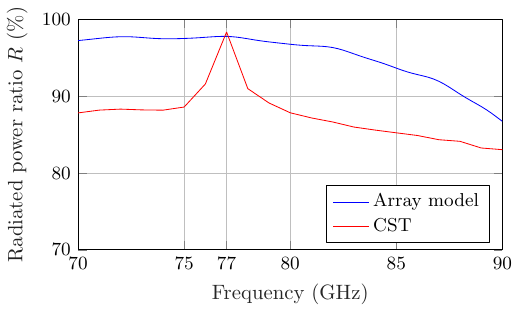}
    \caption{Top unidirectionality $R$ in the design frequency range for the array model given in~\secref{sec:ThArray}, and the CST simulation of the real radiating structures. A high $R\geq \qty{98}{\percent}$ of the power is expected to radiate at the top at $f_{\textrm{bs}}$. The full-wave simulation takes all coupling effects into account, showing a reduced unidirectionality away from the broadside frequency.}
    \label{fig:R_top_RadStr}
\end{figure}

Fig.~\ref{fig:R_top_RadStr} shows the unidirectionality results of the array model and the full-wave simulation of the radiating structures over the design frequency range. The two methods deliver a similar unidirectionality performance at the broadside frequency, with \qty{98}{\percent} of the radiated power being directed to the upper hemisphere. This high unidirectionality performance is present even at lower frequencies in the array model, only tapering down after $f_{\textrm{bs}}$ to reach \qty{87}{\percent} at \qty{90}{\GHz}. The real radiating structures simulated in CST simulations show however that the high unidirectionality is actually achieved only at broadside, with the performance reduced for both lower and higher frequencies, reaching \qty{83}{\percent} as a lowest value. This is because the array model assumes a perfect and simple interference from the isotropic elements scattering, and does not model wave harmonics coupling effects.

\section{Prototype and fabrication}\label{sec:Proto}
With the AD-LWA radiating geometry established, the final design must account for practical constraints associated with fabrication, feed integration, and measurement. These constraints determine the realizable geometry and consequently influence both the simulated and measured antenna performance.

\subsection{3-D printing process}\label{sec:Proto_print}

As a way to demonstrate the feasibility of an innovative manufacturing process with high-performance HF material, the antenna is 3\nobreakdash-D printed with alumina using a Lithoz GmbH CeraFab 7500 3\nobreakdash-D printer at the Institute of Technology for Nanostructures (NST) of the Duisburg-Essen university. The process is defined as Lithography for Ceramic Manufacturing (LCM), where a slurry containing the \ce{Al2O3} particles and a bonding resin is applied on a table with a screen at its bottom part. The screen is illuminated from below, hitting a thin layer of the slurry and hardening it, bonding it to a plate that can move in the $z$-direction (height). The final "green" structure obtained at the end of the print needs to be cleared of the resin agent and hardened (sintered) in an oven with precise heat profiles.\par

Several constraints concerning the possible geometries of the antennas are given:

\begin{outline}
    \1 The maximum printing plate dimensions are \qtyproduct{64 x 40}{\mm}.
    \1 We require a strengthening structure around the antenna to stabilize it mechanically since the structure is very fine.
    \1 The printer screen forces a printing resolution of \qty{25}{\um}. After the sintering process, the green structure hardens and shrinks differently between the planar $xy$ and vertical $z$ directions. Based on the experience at NST, we took the final pixel values $\Delta_{xy}=\qty{20.4}{\um}$ and $\Delta_z=\qty{19.6}{\um}$ --- any parameter of the antenna will need to be a multiple of these two coefficients.
    \1 Additional printing constraints were given in \secref{sec:ThOpenBG} and taken into consideration during the design.
    \1 For the lens design presented below, many constraints linked to the feasibility and quality of the print were taken into account (resolution in $z$ and minimum hole diameter).
\end{outline}

\subsection{Feed and lens structure}\label{sec:Proto_feed}

For the purpose of antenna characterization, the energy input to the antenna will be in the form of a rectangular waveguide (RWG) operating in the $\textrm{TE}_{10}$ mode. An adaptive structure is required to couple the $\textrm{TE}_{10}$ to the DWG $\textrm{TE}_{0}$ mode --- the same triangular design will be used as in~\cite{Francois_EuCAP_04_2026}.\par

Contrary to this reference however, the gratings will not be arranged with circular profiles but will be straight along the width of the antenna. This decision stems from a better compatibility with the printing process, and especially since it results in improved directivity and cross-polarization for frequencies other than broadside. The conversion from circular to planar wave front will be made with the help of a Mikaelian lens~\cite{Mikaelian_ProgOpt_XX_1980}. The design starts with the permittivity $\varepsilon(y)$ profile formula along the $y$ direction (lens width):
\begin{equation}\label{eq:Mik_eps_TEM}
    \varepsilon(y) = \frac{\varepsilon_{\textrm{sub}}}{\mathrm{cosh}^2 \left( \frac{\pi y }{2L_{\mathrm{lens}}} \right)},
\end{equation}
with $L^{\textrm{lens}}$ being the total length of the lens in the $x$ direction. These values are then converted to $\varepsilon_{\textrm{eff}}(y)$ to be compatible with the $\textrm{TE}_{0}$ mode of the DWG, where effective permittivities account for the fields existing outside the substrate. This is achieved by computing the guided wavenumber $\beta_{\varepsilon(y)}^{\mathrm{bs}}$ of the $\textrm{TE}_{0}$ mode for each $\varepsilon(y)$ and computing the effective permittivity as:
\begin{equation}\label{eq:Mik_eps_TE}
    \varepsilon_{\mathrm{eff}}(y) = \left( \frac{\beta_{\varepsilon(y)}^{\mathrm{bs}}}{k_0^{\mathrm{bs}}} \right) ^2,
\end{equation}
with $k_0^{\textrm{bs}}$ the free space wavenumber at $f_\mathrm{bs}$.\par

Next, we need to obtain the $\varepsilon_{\textrm{eff}}(y)$ profile with manufacturing considerations. As presented in~\cite{Zheng_JLT_12_2024}, one can discretize the lens in small unit cells of period $p_{\mathrm{lens}}$ with holes in them. The hole diameters will then control the effective permittivity, which can be obtained with a unit cell simulation with CST and the algorithm described in~\cite{Chen_PRE_07_2004}. Due to the manufacturing constraints mentioned in \secref{sec:Proto_print}, it was not possible to obtain all the required $\varepsilon_{\textrm{eff}}(y)$ values with a fine enough resolution using simply holes in the substrate. An additional design step involved changing also the height of the substrate, such as mentioned in~\cite{Bor_IJMWT_07_2014}, to add a degree of freedom for the available permittivities, since the substrate height also influences the effective permittivity of the $\textrm{TE}_{0}$ mode. 

\begin{figure}[!htbp]
    \centering
    \includegraphics[width=\columnwidth]{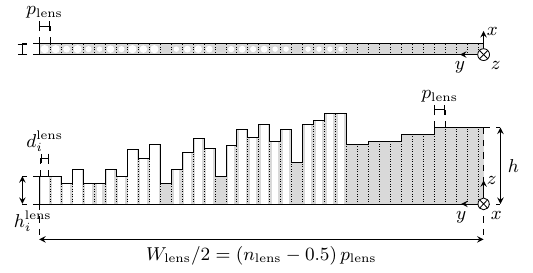}
    \caption{Mikaelian lens bottom and side profile (half left side looking towards the AD-LWA) views. The rightmost vertical dashed line represents the center of the lens, at the feed position with a substrate height corresponding to the AD-LWA height $h$. The parameters $p_{\mathrm{lens}}$, $h_{i}^{\mathrm{lens}}$ and $d_{i}^{\mathrm{lens}}$ (holes diameters represented as white in the figure) allow the permittivity profile of the lens to generate a planar wavefront. The total width $W_{\textrm{lens}}$ of the AD-LWA is derived from the total unit cells of the lens $n_{\mathrm{lens}}$ and their period.}
    \label{fig:dia_lens_prof}
\end{figure}

The final expected printed nominal geometry is given in Fig.~\ref{fig:dia_lens_prof} with half width bottom and side profile views. With this structure, we have $\varepsilon_{\textrm{eff}}^{\textrm{UC}}(y) = f \left( h_{i}^{\mathrm{lens}}; d_{i}^{\mathrm{lens}} \right)$ with $i=[1; \, n_{\mathrm{lens}}]$ the unit cell number along $y$.

\begin{figure}
    \centering
    \captionsetup[subfigure]{oneside,margin={1cm,0cm}}
    \begin{subfigure}{0.47\linewidth}
        \centering
        \includegraphics[height=75.119mm]{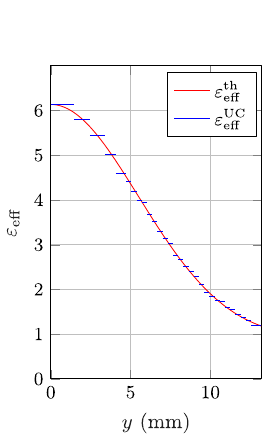}
        \caption{}
        \label{fig:lens_eps_eff_comp}
    \end{subfigure}
    \hspace{0.05cm}
    \begin{subfigure}[b]{0.495\linewidth}
        \centering
        \includegraphics[height=75.119mm]{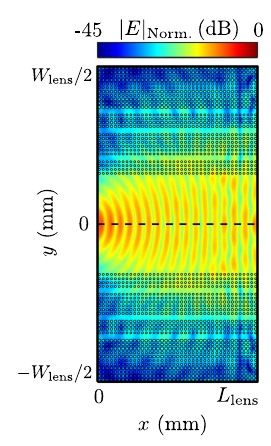}
        \caption{}
        \label{fig:lens_efield_dB}
    \end{subfigure}
    \caption{\protect\subref{fig:lens_eps_eff_comp}~Theoretical (continuous $\varepsilon_\mathrm{eff}^{\mathrm{th}}$) and obtainable (discrete $\varepsilon_\mathrm{eff}^{\mathrm{UC}}$) effective permittivity profiles along a half width Mikaelian lens with substrate permittivity $\varepsilon_\mathrm{sub}=9.2$ (Fig.~\ref{fig:dia_lens_prof}). The maximum effective permittivity at $y=0$ is $\varepsilon_{\textrm{eff}}(0)=6.13$. \protect\subref{fig:lens_efield_dB}~Simulated normalized E-field at $f_\mathrm{bs}$ in the lens at an arbitrary height $z=\qty{0.2}{\mm}$ of the final printed antenna, showing the circular to planar wavefront conversion.}
    \label{fig:lens_data}
\end{figure}

Fig.~\ref{fig:lens_data}\subref{fig:lens_eps_eff_comp} shows the theoretical and obtainable effective permittivity values calculated for a Mikaelian lens of total width $W_{\textrm{lens}}$ and length $L_{\textrm{lens}}$. The resulting conversion of the wavefront from circular to planar is shown in Fig.~\ref{fig:lens_data}\subref{fig:lens_efield_dB}, with the normalized E-field magnitude plotted along the surface of the lens at an arbitrary height $z=\qty{0.2}{\mm}$ (the field being entirely in the substrate).

\subsection{Final prototype and print deviations}\label{sec:Proto_final}

\begin{figure}[!htbp]
  \centering
  \includegraphics[width=\linewidth]{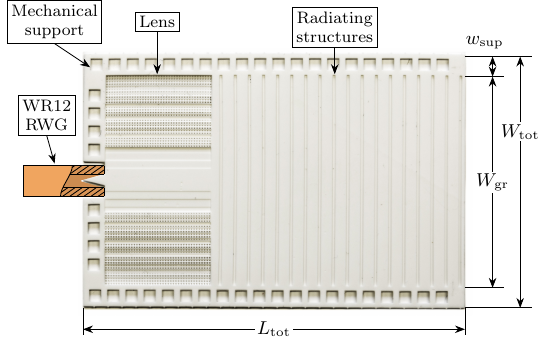}
  \caption{Fabricated prototype (top view) with its different elements and dimensions. The final prototype has the following dimension values: $W_{\textrm{gr}} = W_{\textrm{lens}} = \qty{26.4}{\mm}$, $w_{\textrm{sup}} = \qty{2.75}{\mm}$, $W_{\textrm{tot}} = \qty{31.9}{\mm}$ and $L_{\text{tot}} = \qty{48.3}{\mm}$. We additionally show a WR12 RWG waveguide section to illustrate the feed coupling from the RWG $\textrm{TE}_{10}$ to the DWG $\textrm{TE}_{0}$ modes.}
  \label{fig:prototop_overlay}
\end{figure}

After taking into account all the print constraints and defining the antenna performance in simulation, the final antenna geometry was printed and is shown in Fig.~\ref{fig:prototop_overlay}. 

The nominal design geometrical values of the AD-LWA radiating structures are given in Table~\ref{tab:param_rad_str}. Two radiating structures geometry are given: the theoretical values from the  \secref{sec:Theory} investigation, and the effective nominal structure due to printing constraints. Due to time considerations, there was no redesign of the gratings to accommodate the period $p$ deviation. The lens parameters are given in Table~\ref{tab:param_lens}.  

\begin{table}[!htbp]
    \centering
    \caption{Radiating structures parameters.}
    \begin{tabular}{l||cccc|c}
        \toprule
         & $h$  & $p$ & $l$ & $\delta_x$ & \multirow{2}{*}{$N$} \\
         & \multicolumn{4}{c|}{(\unit{\um})} &  \\
        \midrule
        Theory & 642  & 1760  & 346 & 470 & \multirow{2}{*}{18} \\
        Nominal & 647  & 1773  & 346 & 469 &  \\
        \bottomrule
    \end{tabular}
    \label{tab:param_rad_str}
\end{table}

\begin{table}[!htbp]
\centering
\caption{Lens parameters.}
    \setlength{\tabcolsep}{2.5pt}
    \begin{tabular}{l||ccc|ccc}
        \toprule
         & $p_{\mathrm{lens}}$ & $h_{i}^{\mathrm{lens}}$ & $d_{i}^{\mathrm{lens}}$ & \multirow{2}{*}{$n_{\mathrm{lens}}$}  & \multirow{2}{*}{$W_{\mathrm{lens}}$} & \multirow{2}{*}{$L_{\mathrm{lens}}$} \\
        & \multicolumn{3}{c|}{(\unit{\um})} &  \\
        \midrule
        Nominal & 326 & \qtyrange{176}{764}{} & \qtyrange{163}{244}{} & 41 & $(2n_{\mathrm{lens}}-1) p_{\mathrm{lens}}$ & $p_{\mathrm{lens}} n_{\mathrm{lens}}$  \\
        \bottomrule
    \end{tabular}
    \label{tab:param_lens}
\end{table}

The printed prototype antenna showed final printed geometries deviating from the nominal ones presented in Tables~\ref{tab:param_rad_str} and~\ref{tab:param_lens} (different substrate $h$ and grating heights, lens heights $h_{i}^{\mathrm{lens}}$ and holes diameters $d_{i}^{\mathrm{lens}}$). These deviations were however not deemed to be too critical and the prototype radiation performance was measured, with results presented in the next section.

\section{AD-LWA simulation and measurement results}\label{sec:Res}

In this section we describe the measurement system used to characterize the radiation characteristics of the antenna, and compare the measurement results with simulations. Note that for all the upcoming simulation results, we defined $\tan\delta = 0.0006$ the loss tangent of the alumina material.

\subsection{Measurement setup}
\begin{figure}
    \centering
    \begin{subfigure}{0.49\linewidth}
        \centering
        \includegraphics[width=\linewidth]{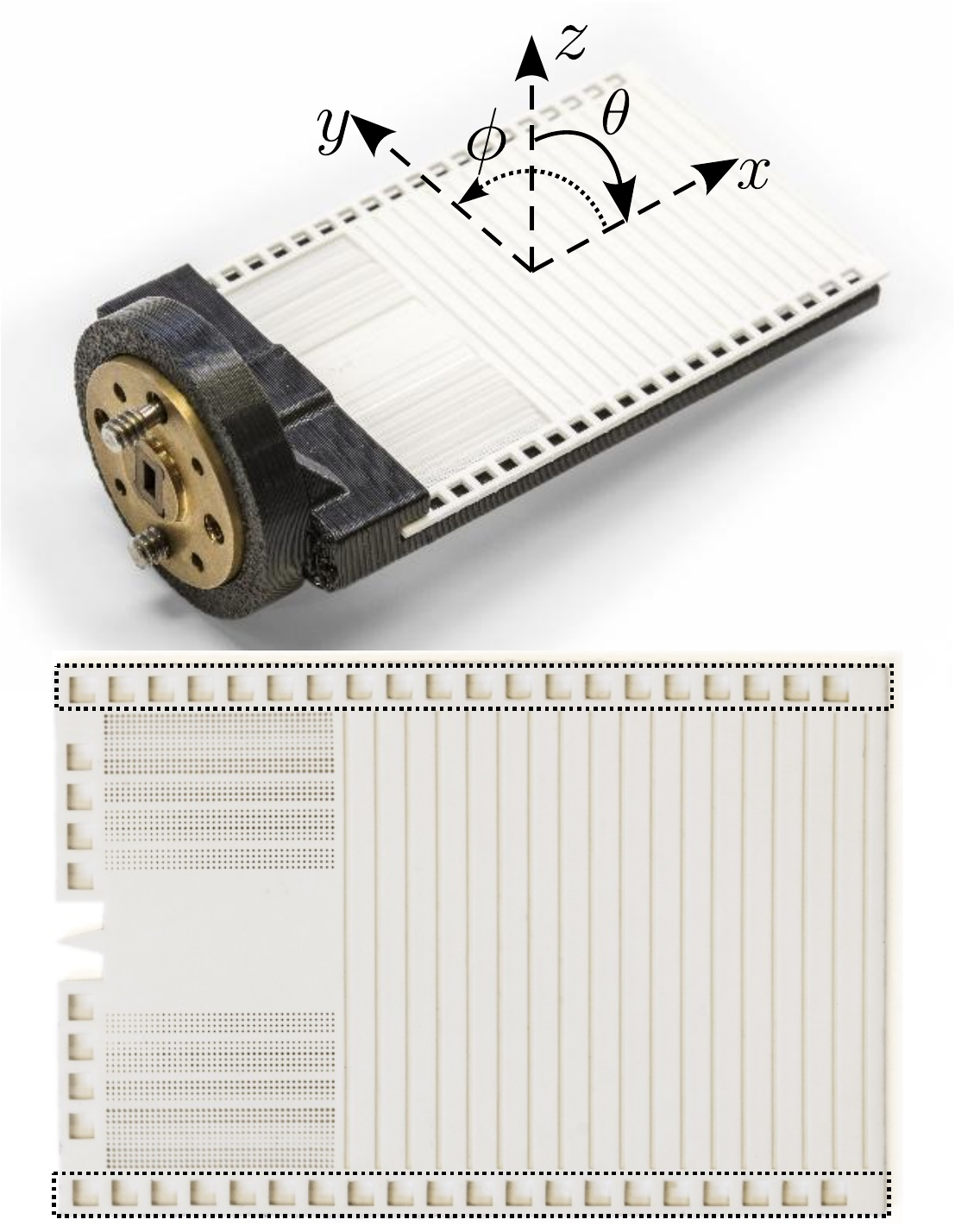}
        \caption{}
        \label{fig:ProtoWSupport}
    \end{subfigure}
    \begin{subfigure}{0.49\linewidth}
        \centering
        \includegraphics[width=\linewidth]{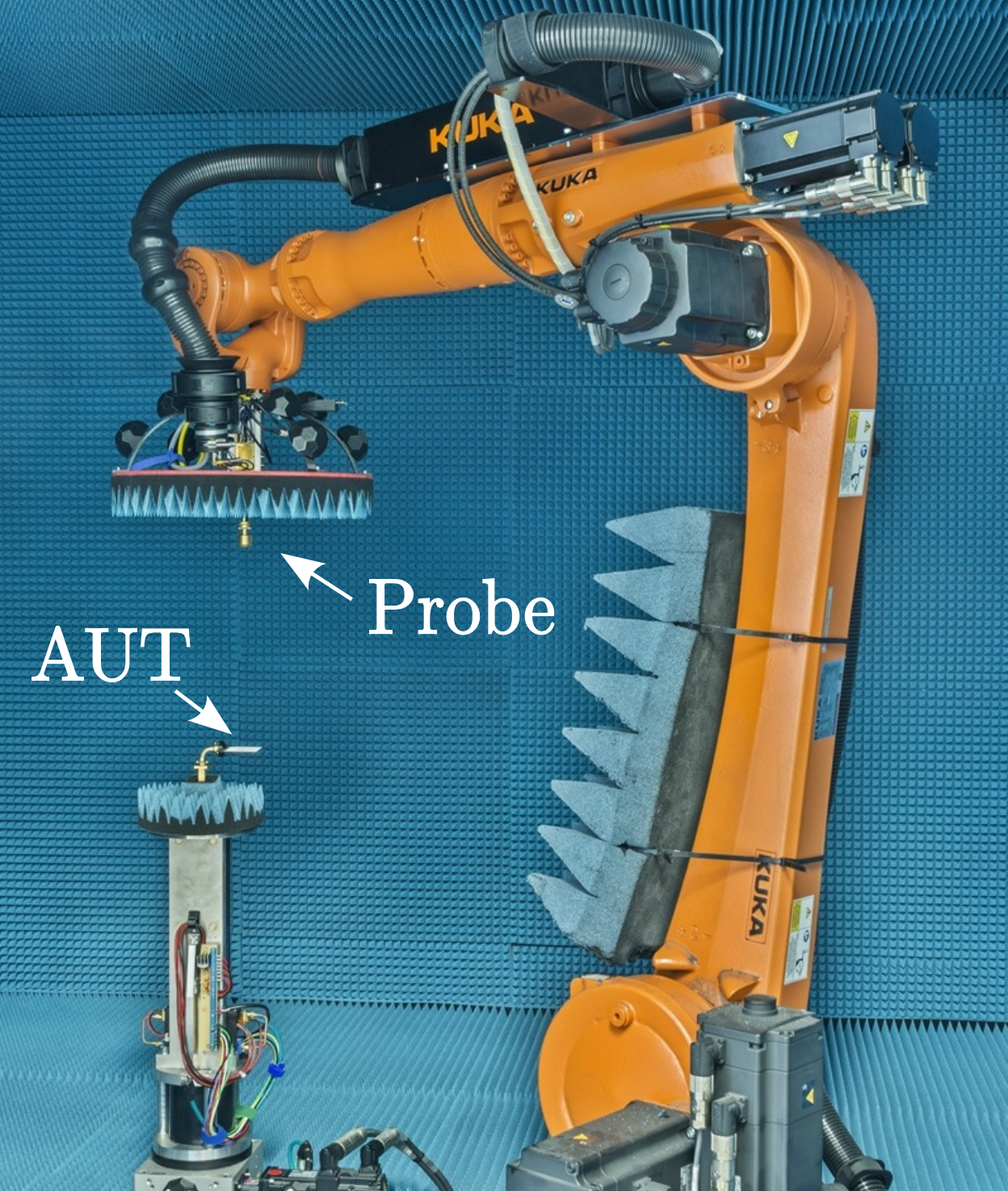}
        \caption{}
        \label{fig:MeasRobot}
    \end{subfigure}
    \caption{Measurement setup for the AD-LWA prototype. \protect\subref{fig:ProtoWSupport}~Final antenna prototype fully assembled, inserted in the WR12 waveguide and mounted on its support. Axes definition and measurement angles for the radiation patterns are shown. Also provided is a bottom view of the antenna showing the support placement, without overlap with the lens and radiating structures. \protect\subref{fig:MeasRobot}~Antenna assembly installed on the IHF robotic measurement system.}
    \label{fig:AntMeas}
\end{figure}

After the antenna was manufactured, a support structure was 3\nobreakdash-D printed to support the antenna horizontally as this will be its mounting position in the measurement system. A WR12 RWG integrated in a flange is inserted in the printed support to provide the connection from the measurement system to the AD-LWA. The whole final AD-LWA assembly ready for measurements is presented in Fig.~\ref{fig:AntMeas}\subref{fig:ProtoWSupport}, with a coordinate system angles conventions described for upcoming plots. We also show a bottom view of the antenna, with the placement of the support structures, which do not overlap with the bottom lens and radiating structures.

The measurements were carried out with the robotic antenna measurement system at the Institute of High Frequency Technology (IHF), RWTH Aachen University. The system comprises a robotic arm for probe positioning and an additional rotation axis mounted on a linear rail for antenna under test (AUT) positioning. The setup shown in Fig.~\ref{fig:AntMeas}\subref{fig:MeasRobot}, including the mounted prototype, is configured for a near-field antenna measurement~\cite{Moch_EuCAP_04_2021}.

While the robot is highly flexible, its displacement constraints limit the measurement to the AUT upper hemisphere (the $z \ge 0$ region), truncating the valid far-field pattern region accordingly. Although this is sufficient for most high-directivity antennas, it would not allow is to validate the back radiation of the prototype. Since the antenna is side-fed, however, it can be rotated by \qty{180}{\degree} about the $x$-axis so that the lower hemisphere can be measured as well. A further advantage of this approach is that the back radiation is not obstructed by the support structure, as is the case for typical positioning systems. However, since the antenna center is offset from the rotation axis, a corresponding offset must be applied to the spherical cut to properly align the robot and the AUT~\cite{Jansen_AMTA_10_2023}. The two near-field measurements, covering the lower and upper hemispheres, are finally combined prior to the near-field-to-far-field transformation to obtain a valid far-field pattern over the full sphere.

This enables not only a characterization of the back-radiation, but also an accurate determination of the antenna directivity. The gain is evaluated by measuring a reference standard-gain horn antenna and applying the established gain-comparison method. In addition to radiation pattern measurements, the system allows the $S_{11}$ of the antenna to be measured within the controlled chamber environment. Absorbers are placed below the AUT to avoid errors caused by reflections from the support structure, which would otherwise arise from the non-negligible back-radiation or, in the flipped configuration, from the main lobe.

\subsection{Reflection coefficient}\label{sec:ResS11}

\begin{figure}[!htbp]
    \centering
    \includegraphics[width=\columnwidth]{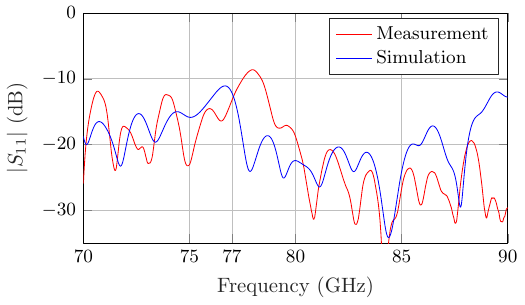}
    \caption{$\abs{S_{11}}$ comparison of the prototype antenna with simulation. The measurement shows the antenna is well matched across the whole frequency range. A frequency shift of the peak $\abs{S_{11}}$ from \qty{77}{\GHz} to \qty{78}{\GHz} shows the discrepancy in material/geometry parameters.}
    \label{fig:S11}
\end{figure}

The measured $\abs{S_{11}}$ of the antenna is presented in Fig.~\ref{fig:S11} and compared to simulation. A good matching is achieved across the whole frequency range with values mostly below \qty{-10}{\dB}. One can observe the $\abs{S_{11}}$ peak shifting from $\approx\qty{77}{\GHz}$ in simulation to a measured $\approx\qty{78}{\GHz}$, indicating a probable discrepancy also for radiation performances, which will be discussed below.

\subsection{Radiation pattern}\label{sec:ResRadPat}

\begin{figure}[!htbp]
    \centering
    \includegraphics[width=\columnwidth]{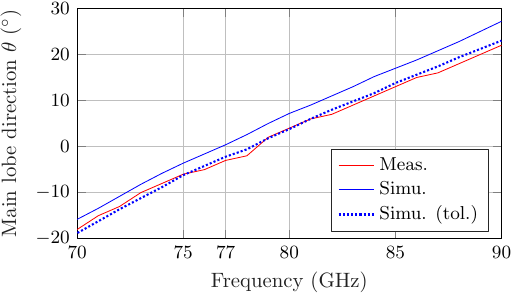}
    \caption{Simulated and measured main lobe directions in the $\phi=\qty{0}{\degree}$ plane. The antenna scans in $\theta$ from \qty{-18}{\degree} to \qty{22}{\degree} over the frequency range. The measured broadside frequency (main lobe direction \qty{0}{\degree}) is around \qty{78}{\GHz}. Dotted line: simulation taking into account printing manufacturing tolerances and material permittivity.}
    \label{fig:MainLobe}
\end{figure}

The first property confirmed by the measured radiation pattern is the beam scanning of the antenna. Fig.~\ref{fig:MainLobe} shows the main lobe direction in the $\phi=\qty{0}{\degree}$ plane with the angle conventions defined in Fig.~\ref{fig:AntMeas}\subref{fig:ProtoWSupport}. The scanning leaky-wave characteristic is clearly visible, with the main lobe steering from \qty{-18}{\degree} to \qty{22}{\degree} over the measured frequency range.\par

There is a systematic deviation between simulation and measurement, resulting in broadside radiation ($\theta=\qty{0}{\degree}$) reached at $\qty{78}{\GHz}$ instead of the design value $f_{\textrm{bs}}=\qty{77}{\GHz}$ --- this is illustrating the frequency shift caused by material and geometrical inaccuracies. We further demonstrate the causes of the frequency shift by performing another simulation of the AD-LWA, implementing measured geometrical values in the modelas mentioned in \secref{sec:Proto_print}, and especially changing the permittivity $\varepsilon_{\textrm{sub}}$ from $\qty{9.2}{}$ to $\qty{8.75}{}$ (this permittivity value was fitting simulation much better with measurement and could be a realistic deviation from the design value taken). In this case the simulation fits measurement much more closely, demonstrating the validity of the design if all manufacturing and material influences can be taken into account in the design phase. These same tolerance parameters will be used in further results below. 

The same shift is observed with the directivity. While the simulation predicts maximum directivity at \qty{77}{\GHz}, the measured maximum occurs at \qty{78}{\GHz} with a measured top directivity of \qty{26}{\dBi}. The agreement remains good across the entire frequency range, with the measured directivity slightly exceeding the simulated one. A sharp dip in the maximum directivity over the lower hemisphere is observed at \qty{78}{\GHz}, again showing an approximate \qty{1}{\GHz} shift in the measurements. Simulations including material tolerances, as indicated above, show even better agreement with the measurement for the bottom-hemisphere directivity. We can also see that the antenna is highly efficient, with measured efficiencies being no worse than $\qty{-1}{\dB}$.

\begin{figure}[!htbp]
    \centering
    \includegraphics[width=\columnwidth]{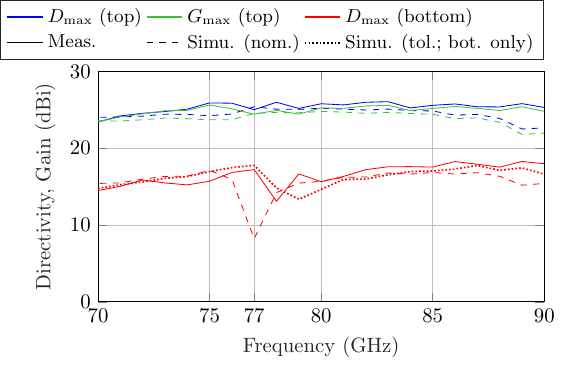}
    \caption{Measured and simulated maximum directivity and gain for the top and bottom hemispheres over frequency. There is a good agreement with simulation for the top hemisphere with a measured top directivity of \qty{26}{\dBi} at broadside (\qty{78}{\GHz}). Dashed lines: simulation with nominal print geometrical values; Dotted line: simulation taking into account printing manufacturing tolerances and material permittivity.}
    \label{fig:Dir}
\end{figure}

\begin{figure}[!htbp]
    \centering
    \includegraphics[width=\columnwidth]{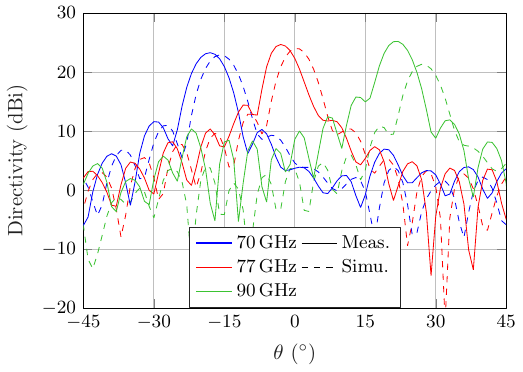}
    \caption{Measured and simulated radiation patterns at \qty{70}{}, \qty{77}{} (design broadside frequency) and \qty{90}{\GHz} in the $\phi=\qty{0}{\degree}$ plane. The discrepancy in the main lobe directions is clearly visible, as synthesized in Fig.~\ref{fig:MainLobe}, even though the main beam shapes are similar.}
    \label{fig:RadPat_nom}
\end{figure}

Finally, representative cuts of the radiation pattern ($\phi=\qty{0}{\degree}$) at selected frequencies (minimum, broadside and maximum) are presented in Fig.~\ref{fig:RadPat_nom}. As expected, the measured main beam direction is offset due to the general frequency shift of the antenna; otherwise, the agreement between the individual patterns is good, except at \qty{90}{\GHz}, where even larger deviations between simulation and measurement are observed, similar to the discussion of the directivity [Fig.~\ref{fig:Dir}].

\begin{figure}[!htbp]
    \centering
    \includegraphics[width=\columnwidth]{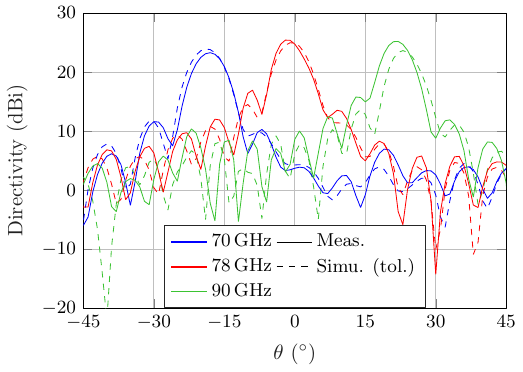}
    \caption{Measurement comparison of the patterns with simulation taking into account printing and material tolerances --- at \qty{70}{}, \qty{78}{} (achieved broadside frequency) and \qty{90}{\GHz} in the $\phi=\qty{0}{\degree}$ plane. One can observe a much better agreement between simulation and measurement in this case, which is promising for future designs. Note however the increased discrepancy at \qty{90}{\GHz}.}
    \label{fig:RadPat_tol}
\end{figure}

These cuts are reproduced with the updated simulation model taking tolerances into account and results match measurement much more closely, as is shown in Fig.~\ref{fig:RadPat_tol}.

\subsection{Unidirectionality}\label{sec:ResUniDir}

The AD-LWA unidirectionality results follow the definition of $R$ given in~\eqref{eq:unidir_R_CST} in~\secref{sec:ThSimRes}. As a first step we investigate the simulated results of the different configurations leading from the theory to the fully defined AD-LWA structure, and the effect on $R$.

\begin{figure}[!htbp]
    \centering
    \includegraphics[width=\columnwidth]{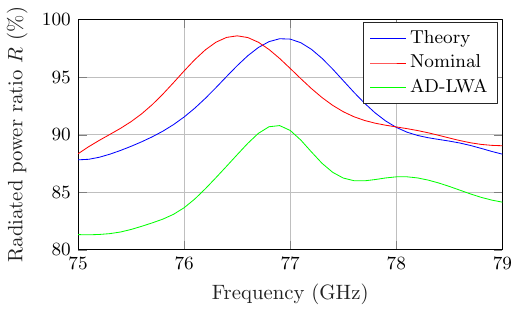}
    \caption{Simulated radiated power ratio $R$ over a narrow frequency range $\left[ f_{\mathrm{bs}}-2; \, f_{\mathrm{bs}}+2 \right]$ near peak unidirectionality, for different radiating structure scenarios. The theoretical grating structures and the nominal ones given in Table~\ref{tab:param_rad_str} were simulated with perfectly matched ports, while the AD-LWA simulation corresponds to the final complete antenna structure. The final AD-LWA unidirectionality shows worse performance, owing mainly to the non-matched condition at the end of the radiating structures.}
    \label{fig:RadPow_BigSc}
\end{figure}

In Fig.~\ref{fig:RadPow_BigSc} we show the unidirectionality parameter $R$ over a narrow frequency range $\left[ f_{\mathrm{bs}}-2; \, f_{\mathrm{bs}}+2 \right]$ near the design broadside frequency. Three cases are presented:

\begin{enumerate}
    \item The radiating structures with the simulation configuration as described in~\secref{sec:ThSimRes} and the geometry described in Table~\ref{tab:param_rad_str} as "Theory".
    \item The same configuration but with the "Nominal" geometry due to printing, showing potential expected performance differences.
    \item The full AD-LWA configuration, with: a real feed+lens configuration, thus having no perfect planar wavefront at the beginning of the radiating structures; and no matched port at the end of the radiating structure. 
\end{enumerate}

\noindent We see that at $f_{\textrm{bs}} = \qty{77}{\GHz}$, the structures with geometry based on the theoretical investigations (blue curve) are effectively radiating more than \qty{98.3}{\percent} of the total radiated power in the upper hemisphere. The final nominal geometry, while having a broadside frequency shifted to \qty{76.5}{\GHz}, shows also a promising performance of $R = \qty{98.5}{\percent}$ (red curve). For the full AD-LWA one can however see that $R = \qty{91.3}{\percent}$ at a broadside frequency close to $f_{\textrm{bs}} = \qty{77}{\GHz}$ (green curve), which is the prototype expected performance. Although this represents a lower unidirectionality performance, one could improve it by further refining the planar wavefront generated by the lens, adjusting the leakage rate of the structures and/or integrating matching structures to reduce back reflections to the antenna.

\begin{figure}[!htbp]
    \centering
    \includegraphics[width=\columnwidth]{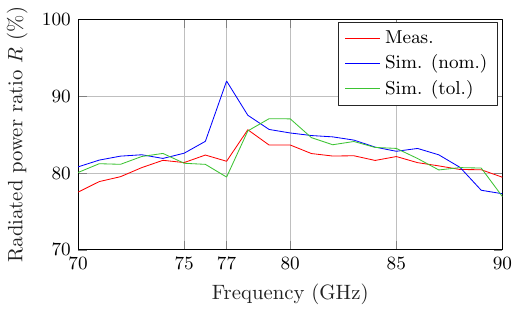}
    \caption{Measured and simulated top radiated power ratio over the whole frequency range. The measured top hemisphere radiation is lower than the simulated one by at most 7 percentage points, while staying mostly above \qty{80}{\percent} for the entire frequency range. The better agreement of the simulation taking into account tolerances can further be observed.}
    \label{fig:RadPowTop}
\end{figure}

Fig.~\ref{fig:RadPowTop} compares the measured and simulated top radiated power ratio of the full AD-LWA across the entire frequency range. The measured upper-hemisphere ratio falls below the simulated one by at most 7 percentage points and remains mostly above \qty{80}{\percent}, confirming the unidirectional behavior of the antenna. The peak observed in the simulation at \qty{77}{\giga\hertz} is not reproduced in the measurement. Instead, a smaller peak appears at \qty{78}{\giga\hertz} with $R = \qty{85.6}{\percent}$, further corroborating the frequency shift discussed in the previous section. This result was reproduced with some fidelity with the tolerance simulation, showing $R$ values at the top close to the measurement.

Even though discrepancies were observed in the prototype measurement compared to the design, the ability to find simulation parameters fitting the measurement is promising for future designs of AD-LWA. Also, an \qty{80}{\percent} radiated power ratio over most of the frequency range is promising for future developments.

\section{Conclusion}\label{sec:conc}
A ground-plane-free all-dielectric leaky-wave antenna (AD-LWA) operating in the frequency range \qtyrange{70}{90}{\GHz} was realized as a monolithic 3\nobreakdash-D-printed alumina radiating structure. The antenna's unit cell consists of two gratings within the slab that are longitudinally shifted to achieve unidirectional radiation and also suppress the broadside open stopband. While the radiation mechanism was explained by simple array theory, a Bloch-wave analysis was used in the design of the shifted gratings and the suppression of the open stopband problem at broadside. Additionally, a novel design of a Mikaelian lens that is printer-compatible is introduced and was integrated to transform the cylindrical wavefront into a planar wavefront of the guided wave.

The fabricated prototype scanned through broadside in a \qtyrange{-18}{22}{\degree} range and achieved a measured realized gain above \qty{23}{\dBi}. Broadside occurred at \qty{78}{\GHz}, approximately \qty{1}{\GHz} above the design frequency. At their respective broadside frequencies, \qty{85.6}{\percent} of the total radiated power was measured in the upper hemisphere, compared with \qty{91.3}{\percent} predicted for the nominal complete antenna model. Simulations incorporating the measured geometric deviations and a fitted relative permittivity of \qty{8.75}{} reproduced the principal frequency shift and directionality trend, highlighting the sensitivity to fabrication tolerances and material properties. These results demonstrate the feasibility of ceramic additive manufacturing for ground-plane-free, all-dielectric LWA radiating structures.

\bibliographystyle{IEEEtran}
\bibliography{IEEEabrv,TAP_2026}
\end{document}

%% file: Z_Settings.tex
\usepackage{balance}

\usepackage[nohyperlinks, nolist]{acronym}

\usepackage{rotating}
\usepackage{algorithmic}
\usepackage{amsmath}
\usepackage{amssymb}
\usepackage{amsfonts}
\usepackage{bm}

\usepackage{physics-patch} 

\usepackage{siunitx}	
\DeclareSIUnit{\dBi}{\mathrm{dBi}}
\AtBeginDocument{\RenewCommandCopy\qty\SI} 

\usepackage{booktabs}
\usepackage{gensymb}
\usepackage{hyperref}

\usepackage{nicefrac}
\usepackage[labelformat=parens]{subcaption}
\usepackage{graphicx}
\usepackage{epstopdf}
\usepackage{multirow}
\usepackage{pgfplots}
\usepackage{textcomp}
\usepackage[version=4]{mhchem}
\usetikzlibrary{arrows.meta, shadings, calc}

\pgfplotsset{compat=newest}
\usepgfplotslibrary{polar}

\newlength\figureheight
\newlength\figurewidth

\usepackage{color}
\usepackage{placeins}
\usepackage{hyperref}
\hypersetup{hidelinks}
\usepackage{booktabs}
\usepackage{tabularray}
\usepackage{colortbl} 
\usepackage[dvipsnames]{xcolor}

\usepackage{pgfplots}

\newcommand{\ieeeticklabel}[1]{%
  \pgfmathparse{abs(#1)<10000}%
  \ifdim\pgfmathresult pt>0.5pt
    \pgfkeys{/pgf/number format/1000 sep=}%
  \else
    \pgfkeys{/pgf/number format/1000 sep={\,}}%
  \fi
  \pgfmathprintnumber{#1}%
}

\pgfplotsset{compat=newest, every axis/.append style={
    xticklabel=\ieeeticklabel{\tick},
    yticklabel=\ieeeticklabel{\tick},
  }}

\usepackage{tikz}
\usetikzlibrary{plotmarks}
\usetikzlibrary{arrows.meta}
\usetikzlibrary{external}
\usetikzlibrary{spy}
\usetikzlibrary{calc}

\usepgfplotslibrary{patchplots}
\usepackage{grffile}
\usepackage{csvsimple}
\usepackage{multirow}
\usepackage{longtable}
\usepackage{listings}
\usepackage{placeins}
\usepackage{tabto}
\usepackage{outlines}

\usepackage{cite}

\newcommand{\secref}[1]{Section~\ref{#1}}

%% file: Tikz/F_simp_legend.tikz

\definecolor{mycolor1}{rgb}{0.23100,0.76600,0.19600}%

\begin{tikzpicture}[font=\small]
\def\sampleLen{4mm}   
\def\midGap{3mm}      
\def\lblGap{0.5mm}       
\def\padX{2mm}         
\def\padY{2mm}       

\coordinate (start) at (0,0);
\draw[color=blue, line width=0.4pt] (start) -- ++(\sampleLen,0) coordinate (samp1);
\node[anchor=west] (lbl1) at ($(samp1)+(\lblGap,0)$) {$f_{\textrm{bs}}-\Delta f$};

\coordinate (samp2start) at ($(lbl1.east)+(\midGap,0)$);
\draw[color=red, line width=0.4pt] (samp2start) -- ++(\sampleLen,0) coordinate (samp2);
\node[anchor=west] (lbl2) at ($(samp2)+(\lblGap,0)$) {$f_{\textrm{bs}}$};

\coordinate (samp3start) at ($(lbl2.east)+(\midGap,0)$);
\draw[color=mycolor1, line width=0.4pt] (samp3start) -- ++(\sampleLen,0) coordinate (samp3);
\node[anchor=west] (lbl3) at ($(samp3)+(\lblGap,0)$) {$f_{\textrm{bs}}+\Delta f$};

\draw ($(start)+(-\padX,-\padY)$) rectangle ($(lbl3.east)+(\padX,\padY)$);
\end{tikzpicture}

%% file: Tikz/Field_harmonics_legend.tikz

\begin{tikzpicture}[font=\small]
\def\sampleLen{4mm}   
\def\midGap{3mm}      
\def\lblGap{0.5mm}       
\def\padX{2mm}         
\def\padY{2mm}       

\coordinate (start) at (0,0);
\draw[color=black, line width=0.4pt] (start) -- ++(\sampleLen,0) coordinate (samp1);
\node[anchor=west] (lbl1) at ($(samp1)+(\lblGap,0)$) {$|E_\mathrm{T}|^2 / \mathrm{max}(|E_\mathrm{T}|^2)$};

\coordinate (samp2start) at ($(lbl1.east)+(\midGap,0)$);
\draw[color=blue, line width=0.4pt] (samp2start) -- ++(\sampleLen,0) coordinate (samp2);
\node[anchor=west] (lbl2) at ($(samp2)+(\lblGap,0)$) {$|E_{+1}|^2\,_\mathrm{Norm.}$};

\coordinate (samp3start) at ($(lbl2.east)+(\midGap,0)$);
\draw[color=red, line width=0.4pt] (samp3start) -- ++(\sampleLen,0) coordinate (samp3);
\node[anchor=west] (lbl3) at ($(samp3)+(\lblGap,0)$) {$|E_{-1}|^2\,_\mathrm{Norm.}$};

\draw ($(start)+(-\padX,-\padY)$) rectangle ($(lbl3.east)+(\padX,\padY)$);
\end{tikzpicture}